\documentclass[preprint,showpacs,preprintnumbers,amsmath,showkeys,amssymb,aps]{revtex4}

\usepackage{graphicx}
\usepackage{dcolumn}
\usepackage{bm}

\begin{document}

\title{Theory of nuclear excitation by electron capture for heavy ions}

\author{Adriana P\'alffy}
\affiliation{Institut f\"ur Theoretische Physik, Justus-Liebig-Universit\"at Giessen,
Heinrich-Buff-Ring 16, 35392 Giessen, Germany}
\email{Adriana-Claudia.Gagyi-Palffy@uni-giessen.de}

\author{Zolt\'an Harman}
\affiliation{Max-Planck-Institut f\"ur Kernphysik,
Saupfercheckweg 1, 69117 Heidelberg, Germany}

\author{Werner Scheid}

\affiliation{Institut f\"ur Theoretische Physik, Justus-Liebig-Universit\"at Giessen,
Heinrich-Buff-Ring 16, 35392 Giessen, Germany}

\date{\today}

\begin{abstract}

We investigate the resonant process of nuclear excitation by electron capture,
in which a continuum electron is captured into a bound state of an ion with the
simultaneous excitation of the nucleus. In order to derive the cross section a
Feshbach projection operator formalism is introduced. Nuclear states 
and transitions are described by a nuclear collective model and making use of
experimental data. Transition rates and total cross sections for NEEC
followed by the radiative decay of the excited nucleus are calculated for
various heavy ion collision systems.

\end{abstract}

\maketitle


\section{Introduction}

In the resonant process of nuclear excitation by electron capture (NEEC) a free
electron is captured into a bound atomic shell with the simultaneous excitation
of the nucleus. It is the nuclear physics analogue of dielectronic recombination
(DR), where a resonant excitation of a bound electron occurs. NEEC is the
time-reversed process of internal conversion (IC).
The excited nucleus can then decay radiatively or by internal conversion.
In the later case, a resonant inelastic electron scattering on the nucleus occurs. 

The NEEC recombination mechanism has been presented for the first time in Ref. \cite{Goldanskii}. 
Several studies have been made concerning NEEC in plasmas
\cite{Goldanskii,Harston} or in solid targets \cite{Cue,Kimball1,Kimball2}. In
\cite{Cue}, cross sections of the process are calculated through a scaling procedure
applied to the results of DR, considering that the two processes differ only in
their excitation mechanisms. Following that, in Ref. \cite{Kimball1}, similar
estimates of NEEC cross sections are obtained for the same nuclei by using
experimental nuclear rather than atomic data. A more theoretical approach is
provided in Ref. \cite{Harston} through an explicit treatment of the electron-nucleus
interaction in the capture process, following the theory used for the
calculation of IC coefficients from Ref. \cite{Band}. In Ref.
\cite{Kimball2}, non-relativistic calculations are presented for the case of
NEEC into bare ions channeling through single crystals. The results are
in disagreement with the previous ones from \cite{Cue, Kimball1}. As NEEC has
not been observed experimentally yet, neither in plasmas nor in the case of 
ions in crystals or colliding with electron targets, the magnitude of its cross section 
is in doubt. Similar
discrepancies exist also in the case of theoretical calculations for the similar 
process of nuclear
excitation by electron transition (NEET). NEET is a fundamental but rare mode of
decay of an excited atomic state in which the energy of atomic excitation is
transferred to the nucleus. This corresponds to the time-reversed bound state internal
conversion.  Unlike NEEC, NEET has been observed experimentally
\cite{Kishimoto}, in the same year in which direct evidence of the bound
internal conversion \cite{Carreyre} has been reported, thus opening  a new period in which experimental precision allows atomic shell models to have 
regard for the internal structure of the nucleus. The permanent development of the
experimental techniques and the enhanced possibilities of preparing bare ions and
electron targets make the experimental observation of NEEC a reasonable goal for
the foreseeable future. Thus theoretical calculations for NEEC occurring in 
scattering measurements are particularly useful, especially in finding
candidate isotopes and transitions suitable for experimental observation. 

In both NEEC and NEET, that are at the border-line between atomic and nuclear
physics, electronic orbital energy is converted directly into nuclear energy.
They offer therefore the possibility to explore the spectral properties of heavy
nuclei through atomic physics experiments. Experimental techniques developed for
scattering studies of electron recombination with atomic ions, e.g. experiments
with stored \cite{Schippers,Wolf} or trapped \cite{Antonio,Knapp} ions, can be
applied to gain information on the nuclear structure of several nuclides which is
hardly accessible by nuclear scattering experiments. Especially, NEEC is
expected to allow the determination of nuclear transition energies, the study of
atomic vacancy effects on nuclear lifetime and population mechanisms of excited
nuclear levels.

In this work we calculate total cross sections for NEEC followed by the
radiative nuclear decay for various transitions in heavy elements. Particular
interest has been payed for collision systems where experimental requirements
for the observation of NEEC are likely to be fulfilled. In order to derive the
cross section formula for the process we extended the Feshbach projector
formalism developed and used for the DR \cite{Zimmermann} to account for the
interaction of electronic and nuclear degrees of freedom. The electric and
magnetic electron-nucleus interactions are considered explicitly and the nucleus
is described with the help of a nuclear collective model \cite{Greiner}. The
dynamics of electrons is governed by the Dirac equation as required in the case
of high-Z elements. This formalism is presented in Section \ref{theory}. The
calculation of the NEEC rates for the electric and magnetic transitions as well
as the nuclear model are described in Section \ref{rates}. Numerical results of
the calculation are given in Section \ref{results}. We conclude with a short
Summary (Section \ref{sum}). The derivation of the magnetic interaction operator
related to Section \ref{theory}, as well as some larger formulas involved in
Section \ref{rates} are given in the Appendix. In this work atomic units have
been used unless otherwise mentioned.


\section{\label{theory} Theory of nuclear excitation by electron capture}

In this section we derive the total cross section formula for the NEEC process
followed by the radiative decay of the excited nucleus. We consider that the
electron is captured into the bound state in the Coulomb field of a bare nucleus
and that only the nucleus is decaying by emitting a photon. However, the
derivation of the cross section can be extended to the many-electron case in a 
straightforward way and
also an additional radiative decay of the electron can be treated by applying
the formalism.

\subsection{Decomposition of the Fock space by means of projection operators}

The initial state $|\Psi_i\rangle$ of the system consisting of the nucleus in
its ground state, the free electron, and the vacuum state of the electromagnetic
field can be written as a direct product of the nuclear, electronic, and
photonic state vectors:
\begin{equation}\label{eq:}
| \Psi_i \rangle = | N , \vec{p} m_s , 0 \rangle \equiv
| N \rangle \otimes | \vec{p} m_s \rangle \otimes |0 \rangle \,.
\end{equation}
Here, $\vec{p}$ is the asymptotic momentum of the electron, $m_s$ its spin
projection, and $|N\rangle$ the nuclear ground state. The state $|\Psi_d\rangle$
formed by the resonant capture has the form
\begin{equation}\label{eq:dstate}
| \Psi_d \rangle = | N^*, n_d\kappa_d m_d ,0 \rangle \equiv 
| N^* \rangle \otimes | n_d\kappa_d m_d \rangle \otimes |0 \rangle \,,
\end{equation}
with $n_d$, $\kappa_d$, and $m_d$ being the principal quantum number, Dirac
angular momentum, and magnetic quantum numbers of the bound one-electron state,
respectively. The excited nuclear state is denoted by $|N^*\rangle$. The final
state $|\Psi_f\rangle$ of the NEEC process contains a photon with the wave
number $\vec{k}$ and the transversal polarization $\sigma=1,2$, and the nucleus
which is again in its ground state $|N\rangle$:
\begin{equation}
| \Psi_f \rangle = | N , n_d\kappa_d m_d , \vec{k}\sigma \rangle \equiv 
| N \rangle \otimes | n_d\kappa_d m_d \rangle \otimes |\vec{k}\sigma \rangle \ ,
\end{equation}
\begin{equation}
|\vec{k}\sigma \rangle = a^{\dagger}_{\vec{k}\sigma}| 0\rangle \ .
\end{equation}
Here, the $a^{\dagger}_{\vec{k}\sigma}$ is a photon creation operator. The corresponding
conjugate annihilation operator is denoted by $a_{\vec{k}\sigma}$.

To clearly separate these states in the perturbative expansion of the transition
operator, we introduce operators projecting onto the individual subspaces.
Characterizing the state of the electron in the positive part of the continuous 
spectrum by the energy $\varepsilon$ rather than
the momentum of the free electron, we write the projector $P$ belonging to the
first type of subspace as
\begin{equation}\label{eq:pspectral}
P = \int d\varepsilon \sum_{\alpha} | \alpha \varepsilon \rangle \langle \alpha \varepsilon |  \,.
\end{equation}
For brevity we introduce the multi-index $\alpha$ to stand for all discrete
quantum numbers of the total system. The projection operator of the subspace
spanned by states of the type (\ref{eq:dstate}) is written as
\begin{equation}\label{eq:qspectral}
Q=\sum_q | q \rangle \langle q|, 
\end{equation}
with the cumulative index $q$ introduced again to summarize all discrete quantum
numbers describing the bound electron and the excited nucleus. The subspace of
the state vectors containing one transverse photon is associated with the
projection operator
\begin{equation}\label{eq:rspectral}
R = \sum_q \sum_{\vec{k}\sigma} a^{\dagger}_{\vec{k}\sigma} | q \rangle \langle q|  a_{\vec{k}\sigma} \,.
\end{equation}
Assuming corrections due to two- or more-photon states \cite{Zakowicz} and due to the
presence of the negative electronic continuum
to be negligible, we postulate the following completeness relation:
\begin{equation}\label{eq:completeness}
P+Q+R ={\bf{1}}\ ,
\end{equation}
where  ${\bf{1}}$ is the unity operator of the Fock space.

\subsection{The total Hamiltonian of the system}

The total Hamiltonian operator for the system consisting of the nucleus, the
electron, and the radiation field can be written as
\begin{equation}\label{eq:totalh}
H = H_n + H_e + H_r + H_{en} + H_{er} + H_{nr} \,.
\end{equation}
The Hamiltonian of the nucleus $H_n$ is written in terms of the nuclear
collective model \cite{Greiner} by using creation and annihilation operators of
the collective modes, $B^{\dagger}_{\lambda\mu}$ and $B_{\lambda\mu}$,
\begin{equation}\label{eq:nuclearh}
H_n=\sum_{\lambda\mu}\Omega_{\lambda} B^{\dagger}_{\lambda\mu}B_{\lambda\mu}\ ,
\end{equation}
where $\Omega_{\lambda}$ are the phonon frequencies. The Dirac Hamiltonian of
the free electron is given by
\begin{equation}
H_e = c\vec\alpha\cdot\vec p+(\beta-1)c^2\,,
\end{equation}
whilst the pure quantized radiation field is described by 
\begin{equation}
H_r = \sum_{\vec{k}\sigma} \omega_k a^{\dagger}_{\vec{k}\sigma} a_{\vec{k}\sigma} \,.
\end{equation}
Interactions between the three subsystems are described by the three remaining
Hamiltonians in eq.\ (\ref{eq:totalh}). We adopt the Coulomb gauge for the
electron-nucleus interaction because it allows the separation of the dominant
Coulomb attraction between the electronic and the nuclear degrees of freedom,
\begin{equation}\label{eq:coulomb}
H_{en}=\int d^3r_n\frac{\rho_n(\vec{r}_n)}{|\vec{r}_e-\vec{r}_n|}\ .
\end{equation} 
In eq.\ (\ref{eq:coulomb}), $\rho_n(\vec{r}_n)$ is the nuclear charge density and
the integration is performed over the whole nuclear volume. The interaction of
the electron with the transverse photon field quantized in unit volume is given by
\begin{equation}\label{eq:radhamilton}
H_{er} = \sum_{\vec k\sigma} \sqrt{2\pi c^2 \over \omega_k} \vec\alpha \left(
\vec\epsilon_{\vec{k}\sigma} e^{i\vec{k}\cdot\vec{r}_e} a_{\vec{k}\sigma} + h.c.\right) \,.
\end{equation}
Here,  $\vec\alpha$ is the vector of the Dirac
$\alpha$ matrices $(\alpha_x,\alpha_y,\alpha_z)$, and
$\vec{\epsilon}_{\vec{k}\sigma}$ is the polarization vector of the photons.
Similarly, the interaction of the nucleus with the electromagnetic field is
given by the Hamiltonian
\begin{equation}
H_{nr} = - \sum_{\vec k\sigma} \int d^3r_n\vec{j}_n(\vec{r}_n)  \sqrt{\frac{2\pi}{\omega_k}}
\left(\vec\epsilon_{\vec{k}\sigma} e^{i\vec{k}\cdot\vec{r}_n} a_{\vec{k}\sigma} + h.c.\right)
\ ,
\end{equation}
where  $\vec{j}_n(\vec{r}_n)$ is the nuclear current.

Using the projection operators we can separate the Hamiltonian as follows:
\begin{equation}
H = H_0 + V
\end{equation}
with
\begin{equation} \label{eq:h_0}
H_0 = PHP + QHQ + RHR\ , 
\end{equation}
\begin{eqnarray} 
V\equiv H - H_0 = PHQ &+& QHP + PHR+RHP
\nonumber\\
&+&  RHQ + QHR \,. 
\end{eqnarray}
In this way the effect of the nuclear potential on bound and continuum electron 
states is included to all orders.
The individual terms in the perturbation operator describe transitions between
the different subspaces. For example, $PHQ$ describes the transition of the
bound electron to the continuum, i.e. IC, and $QHP$ accounts for the
time-reversed process of IC, namely, NEEC. $PHR$ and $RHP$ are the lowest-order
operators for photoionization and radiative recombination, respectively. $QHR$
and $RHQ$ account for the radiative excitation of the nucleus and radiative
decay of the nucleus or of the electron in a bound state.

\subsection{\label{pertexp} Perturbation expansion of the transition operator}

The transition operator is defined as
\begin{equation}
T(z) = V + VG(z)V \,,
\end{equation}
where  $G(z)$ is the Green operator of the system given by
\begin{equation}
G(z)= (z-H)^{-1} \,.
\end{equation}
Here, $z$ is a complex energy variable. The cross section for a process can be
expressed by the transition operator as follows: 
\begin{equation}\label{eq:tsigma}
\frac{d\sigma_{i \to f}}{d\Omega_k}(E) = \frac{2\pi}{F_i}
\lim_{\epsilon \to 0+}  |\langle \Psi_f|T(E+i\epsilon) |\Psi_i\rangle|^2 \rho_f \ ,
\end{equation}
with the $\Psi_f$ and $\Psi_i$ as final and initial eigenstates of $H_0$,
respectively. This cross section is differential with respect to the angle
$\Omega_k$ of the photon emitted in the process. $F_i$ denotes the flux of the
incoming electrons, and $\rho_f$ the density of the final photonic states. 

We use the Lippmann-Schwinger equation to write the perturbation series for $T(z)$
in powers of $V$ with the Green function $G_0(z)$ of the unperturbed Hamiltonian $H_0$:
\begin{equation}
T(z) = V + VG_0(z)V + VG_0(z)VG_0(z)V + \dots \,.
\end{equation}
Since the initial state of the NEEC process is by definition an eigenstate of
$P$, and the final state is an eigenstate of $R$, we consider the projection
$RTP$ of the transition operator:
\begin{eqnarray}\label{eq:rtp}
RTP = RVP &+& RVG_0VP + RVG_0VG_0VP 
\nonumber \\
&+&RVG_0VG_0VG_0VP + \dots  
\end{eqnarray}
Here and in the following we omit the argument $z$. The first term in eq.\ (\ref{eq:rtp}) does not contribute to the NEEC process. After a further analysis
of the second term $RVG_0VP= RH_{nr}QG_0QH_{en}P$ and inserting the spectral
resolution (\ref{eq:qspectral}) of $Q$ in the second order in $V$ we arrive to
\begin{equation}\label{eq:rtplowest}
\langle \Psi_f |RTP| \Psi_i \rangle =
\sum_q
\frac{\langle N , n_d\kappa_d m_d , \vec{k}\sigma  | H_{nr} |q \rangle\langle q | H_{en}
| N , \vec{p} m_s , 0 \rangle }{z-E_q^0} \ .
\end{equation}
The energy $E_q^0$ denotes the unperturbed eigenvalue of the state $|q\rangle$.
If we  continue analyzing the perturbation
expansion (\ref{eq:rtp}), the term of third order in $V$ can be written as
\begin{eqnarray}
RVG_0VG_0VP =&& RH_{er}PG_0PH_{en}QG_0QH_{en}P \nonumber\\
            &&+ RH_{er}PG_0PH_{er}RG_0RH_{er}P \nonumber\\
	&&+ RHQG_0QHRG_0RH_{er}P\ . 
\end{eqnarray}
The first two terms do not contribute to the cross section of the considered
NEEC process. The last term is decomposed as
\begin{eqnarray}
\label{eq:transverse}
RHQG_0QHRG_0RH_{er}P =&& RH_{er}QG_0QH_{er}RG_0RH_{er}P \nonumber\\
		&&+ RH_{er}QG_0QH_{nr}RG_0RH_{er}P \nonumber\\
                     &&+RH_{nr}QG_0QH_{er}RG_0RH_{er}P \nonumber\\
&&+ RH_{nr}QG_0QH_{nr}RG_0RH_{er}P \ . 
\end{eqnarray}
Here, the first two terms are not considered, as they describe recombination by
radiative decay of the electron.  The process incorporated in the third term of
(\ref{eq:transverse}) is not possible. The remaining last term,
$QH_{nr}RG_0RH_{er}P$, accounts for the capture of the free electron by
exchanging a virtual transverse photon with the nucleus. As we show in Appendix
\ref{A}, it can be approximated by $QH_{magn}P$, where
\begin{equation}
H_{magn} = - \frac{1}{c} \vec{\alpha} \int d^3r_n \frac{\vec{j}_n(\vec{r}_n)}{|\vec{r}-\vec{r}_n|}
=-\vec{\alpha}\cdot\vec{A}(\vec{r})
\label{hmagn}
\end{equation}
is  the  magnetic interaction Hamiltonian.

We continue the expansion (\ref{eq:rtp}) of the $T$ operator and only consider
the terms that contain $QH_{en}P$ as the first step and $RH_{nr}Q$ as the final
step. The contribution of order $V^4$ can be decomposed as
\begin{eqnarray}\label{eq:v4}
&&RVG_0VG_0VG_0VP 
= RH_{nr}QG_0QH_{ne}PG_0PH_{ne}QG_0QH_{ne}P \\   \nonumber
&&+ RH_{nr}QG_0Q(H_{er} + H_{nr})RG_0R(H_{er} + H_{nr})QG_0QH_{ne}P \,.  \nonumber
\end{eqnarray}
We rewrite the first term as
\begin{eqnarray}
&&RH_{nr}QG_0QH_{ne}PG_0PH_{ne}QG_0QH_{ne}P = \\
&&\sum_{q,q'} RH_{nr}QG_0 |q\rangle \langle q|
H_{ne}PG_0PH_{ne}  |q'\rangle\langle q'| G_0QH_{ne}P \,, \nonumber
\end{eqnarray}
and consider the diagonal matrix element
\begin{equation}
\langle q| H_{ne}PG_0PH_{ne}  |q\rangle = \int d\varepsilon \sum_{\alpha}
\frac{\langle q| H_{ne} |\alpha \varepsilon\rangle \langle \alpha \varepsilon| H_{ne} |q\rangle}{z-E^0} \,,
\end{equation}
with $E^0$ defined by $H_0 |\alpha E^0 \rangle = E^0 |\alpha E^0 \rangle$. Using the equality
\begin{equation}
\lim_{\epsilon \to 0_+} \frac{1}{x+i\epsilon} =  \mathcal{P} \left(\frac{1}{x}\right) -i\pi \delta(x) \,,
\end{equation}
it can be further decomposed into 
\begin{equation}
\int d\varepsilon \sum_{\alpha} \frac{\langle q| H_{ne} |\alpha \varepsilon\rangle
\langle \alpha \varepsilon| H_{ne} |q\rangle}{z-E^0} = 
\Delta E_q^{\rm{NP}} - \frac{i}{2} \Gamma_q^{\rm{IC}} \ ,
\end{equation}
\begin{equation} 
\Delta E_q^{\rm{NP}} \equiv \mathcal{P} \int d\varepsilon \sum_{\alpha} \frac{\langle q| H_{ne} |\alpha \varepsilon\rangle
\langle \alpha \varepsilon| H_{ne} |q\rangle}{z-E^0}  \ ,
\end{equation}
\begin{equation}
\Gamma_q^{\rm{IC}}  \equiv i\pi \sum_{\alpha} \left| \langle q| H_{ne} |\alpha E^0\rangle \right|^2 \ .
\end{equation}
The notation $\Delta E_q^{\rm{NP}}$ was introduced to denote the Coulomb nuclear
polarization correction to the energy of the state $q$ and $\Gamma_q^{\rm{IC}}$
for its internal conversion width. $\mathcal{P}$ denotes the principal value of
the integral. 

In a similar manner, the second term of (\ref{eq:v4}) can be analyzed. It can be
separated into the following four parts:
\begin{eqnarray}\label{eq:transversev4}
  RH_{nr}QG_0Q(H_{er} &+& H_{nr})RG_0R(H_{er} + H_{nr})QG_0QH_{ne}P = \\
  && RH_{nr}QG_0QH_{er}RG_0RH_{er}QG_0QH_{ne}P \nonumber \\
 &&+RH_{nr}QG_0QH_{nr}RG_0RH_{nr}QG_0QH_{ne}P \nonumber \\
 &&+RH_{nr}QG_0QH_{er}RG_0RH_{nr}QG_0QH_{ne}P \nonumber\\
 &&+RH_{nr}QG_0QH_{nr}RG_0RH_{er}QG_0QH_{ne}P \,. \nonumber
\end{eqnarray}
The first term describes the emission and reabsorption of a photon by the
electron recombined into the bound state. Its diagonal matrix element possesses
a real and an imaginary part:
\begin{equation}
\langle q|H_{er}RG_0RH_{er}|q\rangle = \Delta E_q^{\rm{SE}} - \frac{i}{2} \Gamma_q^{\rm{e,rad}} \,.
\end{equation}
$\Delta E_q^{\rm{SE}}$ describes the one-loop self energy correction to the
bound state energy of the electron. The imaginary part is the radiative decay
rate of the electronic state and vanishes in the case of electron capture into
the ground state of the ion. The second term in (\ref{eq:transversev4})
describes the emission and a subsequent reabsorption of a virtual photon by the
nucleus, and its diagonal matrix element reads
\begin{equation}
\langle q|H_{nr}RG_0RH_{nr}|q\rangle = \Delta E_q^{\rm{NSE}} - \frac{i}{2} \Gamma_q^{\rm{n,rad}} \,.
\end{equation}
Here, $\Delta E_q^{\rm{NSE}}$ is the nuclear self energy correction to the
energy of the ion, and $\Gamma_q^{\rm{n,rad}}$ stands for the radiative decay
width of the nucleus in the state $q$. The last two terms of
(\ref{eq:transversev4}) incorporate corrections to the
intermediate state energy due to the exchange of a virtual transverse photon
between the electronic and the nuclear currents. These corrections are neglected
in our treatment as they are expected to be far less than the overall accuracy
of experimental nuclear excitation energies.

Continuing the expansion (\ref{eq:rtp}) of the $T$ operator, the matrix element
of the intermediate state Green operator in fourth order contains terms of the
form
\begin{equation}\label{eq:qq'}
\langle q|H_{i}G_0H_{i}G_0H_{i}G_0H_{i}|q\rangle = \sum_{q'} 
\frac{\langle q|H_{i}G_0H_{i}|q'\rangle \langle q'|H_{i}G_0H_{i}|q\rangle}
{z-E_{q'}} \,,
\end{equation}
where the label $i$ stands for $ne$, $er$ and $nr$. We adopt the so-called
isolated resonances approximation by taking only the diagonal matrix elements
into account, i.e.  we set $q=q'$ in (\ref{eq:qq'}). This approximation is valid
if the distance between neighboring resonances is large with respect to their
total natural widths, which is the case in all systems we studied. Higher-order
terms can be summed then as a geometric progression
\begin{equation}
\frac{1}{z-E_q^0} \sum_{k=0}^{\infty} x^k = \frac{1}{z-E_q^0} \frac{1}{1-x} 
\end{equation}
with the dimensionless quotient
\begin{eqnarray}
x = \frac{1}{z-E_q^0} &&\Big(
\langle q| H_{ne}PG_0PH_{ne}|q\rangle \nonumber  \\
&&+ \langle q| H_{er}RG_0RH_{er}  |q\rangle
+ \langle q| H_{nr}RG_0RH_{nr}  |q\rangle
\Big) \,,
\end{eqnarray}
resulting in
\begin{equation}
\frac{1}{z-E_q^0 - \Delta E_q^{\rm{NP}} - \Delta E_q^{\rm{SE}} - \Delta E_q^{\rm{NSE}}
+ \frac{i}{2} \Gamma_q^{\rm{IC}} + \frac{i}{2} \Gamma_q^{\rm{n,rad}}} \,.
\end{equation}

Thus, the infinite perturbation expansion introduces energy corrections and
widths into the energy denominator of the lowest order amplitude
(\ref{eq:rtplowest}).  The final expression for the transition amplitude of
NEEC into states denoted by $d$ and followed by radiative nuclear decay is then
\begin{eqnarray}
&&\langle \Psi_f |RT(z)P|\Psi_i \rangle =\nonumber \\ 
&&\sum_{d}
\frac{\langle N , n_d\kappa_d m_d , \vec{k}\sigma  | H_{nr} |\Psi_d \rangle
\langle \Psi_d| H_{en}+H_{magn} | N , \vec{p} m_s , 0 \rangle}
{z-E_d^0 - \Delta E_d + \frac{i}{2} \Gamma_d} \,.
\end{eqnarray}

Here we introduce the notation $\Delta E_d = \Delta E_d^{\rm{NP}} + \Delta
E_d^{\rm{SE}} + \Delta E_d^{\rm{NSE}}$ for the energy correction and $\Gamma_d =
\Gamma_d^{\rm{IC}} + \Gamma_d^{\rm{n,rad}}$ for the total natural width of the
excited state $|d\rangle=|N^*,n_d\kappa_d m_d, 0\rangle$.

\subsection{Differential and total cross sections for NEEC}

Equation (\ref{eq:tsigma}) gives the differential cross section in terms of the
matrix element of the projected $T$-operator. Neglecting the interference of
neighboring resonances, and taking into account only a single state $d$, with
the corresponding magnetic substates, the cross section for a given reaction
pathway $i \to d \to f$ is
\begin{eqnarray}
\frac{d\sigma_{i \to d \to f}}{d\Omega_k}(E) &=& 
\frac{2\pi}{F_i}
\left|\langle N_fI_fM_{I_f} , n_d\kappa_d m_d , \vec{k}\sigma  | H_{nr} | N^*_dI_dM_{I_d},n_d\kappa_d m_d, 0\rangle\right|^2
\nonumber \\ &\times&
\frac{\left|\langle N^*_dI_dM_{I_d},n_d\kappa_d m_d, 0| H_{en}+H_{magn} | N_iI_iM_{I_i} , \vec{p} m_s , 0 \rangle\right|^2}
{(E-E_d^0 - \Delta E_d)^2 + \frac{1}{4} \Gamma_d^2}
 \rho_f \,.
\end{eqnarray}
$N_i$ as well as $N_f$ represent the nucleus in the ground state, while $N^*_d$
stands for the intermediate excited nuclear state. The angular momentum $I$ and
its projection $M_I$ are used to denote the nuclear states.  One has to perform
an average over the magnetic substates of the initial state of the system and a
summation over the final states if these are not resolved in a NEEC experiment.
The total cross section is calculated by integrating over the solid angle
$\Omega_k$ of the photon emission and averaging over the direction of the
electron:
\begin{eqnarray}
\sigma_{i \to d \to f}(E) &=& \frac{2\pi}{F_i} 
\sum_{M_{I_f} \sigma} \sum_{M_{I_d}m_d} 
\frac{1}{2(2I_i +1)} \sum_{M_{I_i} m_s} 
\frac{1}{2I_d +1}\nonumber\sum_{M'_{I_d}} \frac{1}{4\pi} \int d\Omega_p\\
&\times& \int d\Omega_k 
\frac{|\langle N_f I_f M_{I_f}, n_d\kappa_d m_d , \vec{k}\sigma  | H_{nr} |
N^*_d I_d M_{I_d},n_d\kappa_d m_d,0\rangle|^2}
{(E-E_d)^2 + \frac{1}{4} \Gamma_d^2} 
\nonumber \\ 
&\times&
|\langle N^*_d I_d M_{I_d},n_d\kappa_d m_d, 0| H_{en}+H_{magn} |  N_i I_i M_{I_i} , \vec{p} m_s , 0 \rangle|^2
\rho_f \ . 
\end{eqnarray}

We denote the corrected energy of the intermediate state by $E_d=E_d^0 + \Delta
E_d$. By introducing the notation
\begin{eqnarray}
Y_n^{i \to d}&=&\frac{2\pi}{2(2I_i+1)}\sum_{{M_{I_i}} m_s}\sum_{M_{I_d} m_d}
\nonumber \\
&\times&
\int d\Omega_p
|\langle N^*_dI_dM_{I_d},n_d\kappa_d m_d,0|H_{en}+H_{magn}|N_iI_iM_{I_i},\vec{p}m_s,0\rangle|^2\rho_i
\label{Yrate}
\end{eqnarray}
for the electron capture rate,
\begin{eqnarray}
A_r^{d \to f} &=& \frac{2\pi}{2I_d + 1} \sum_{M_{I_f} \sigma} \sum_{M_{I_d}} \nonumber\\ 
&\times&
\int d\Omega_k 
|\langle N_f I_f M_{I_f}, n_d\kappa_d m_d , \vec{k}\sigma | H_{nr} | N^*_d I_d M_{I_d},n_d\kappa_d m_d, 0\rangle |^2\rho_f  
\end{eqnarray}
for the radiative transition rate, and 
\begin{equation}
L_d(E-E_d) = \frac{\Gamma_d / 2\pi}{(E-E_d)^2 + \frac{1}{4} \Gamma_d^2}
\end{equation}
for the normalized Lorentz profile and taking into account the relation $F_i\rho_i=p^2/(2\pi)^3$, 
the cross section formula can be written in the condensed form
\begin{equation}
\sigma_{i \to d \to f}(E) = \frac{2\pi^2}{p^2}
\frac{A_r^{d \to f} Y_n^{i \to d}}{\Gamma_d} L_d(E-E_d) \,.
\end{equation}

Determining the total cross section of the studied process requires the
calculation of the transition rates $Y_n$ and $A_r$, and the initial and final
state energies. In the actual calculations we neglect the additional corrections
$\Delta E_d^{\rm{NP}}$ and $\Delta E_d^{\rm{NSE}}$. The integration of the cross
section over the continuum electron energy gives the resonance strength $S_d$
for a given recombined state $d$,
\begin{equation}
S_d=\int dE\ \frac{2\pi^2}{p^2}
\frac{A_r^{d \to f} Y_n^{i \to d}}{\Gamma_d} L_d(E-E_d) \ .
\end{equation}
The natural width $\Gamma_d$ of the nuclear excited state is of the order of
$10^{-6}$ eV. In this interval the values of $p^2$ as well as of the NEEC rate
$Y_n$ can be considered constant. As the Lorentz profile is normalized to unity,
\begin{equation}
\int dE\ L_d(E-E_d)=1 \ ,
\end{equation}
the resonance strength can be written as
\begin{equation}
S_d=\frac{2\pi^2}{p^2}
\frac{A_r^{d \to f} Y_n^{i \to d}}{\Gamma_d}  \ .
\end{equation}


\section{\label{rates} Rates for electric and magnetic transitions}

In order to calculate the NEEC rate we have considered the matrix element of the
electric and magnetic interactions between the electron and the nucleus. We
write the wave function of the system as the product wave function of the
electronic and nuclear states,
\begin{equation}
H_{fi}= \langle N^*I_d M_{I_d} |\langle n_d\kappa_d m_d|H_{en}+H_{magn}|\vec{p} m_s \rangle |N I_i M_{I_i} \rangle\ .
\end{equation}
The initial state continuum electronic wave function is given through a partial
wave expansion \cite{Eichler},
\begin{equation}
|\vec{p} m_s\rangle=\sum_{\kappa m}i^l e^{i\Delta_{\kappa}}\sum_{m_l}Y_{l m_l}^*(\Omega_e)
C\left(l\ \frac{1}{2}\ j;m_l\ m_s \ m\right)| \varepsilon\kappa m\rangle\ ,
\end{equation}
where $\varepsilon$ is the energy of the continuum electron measured from the
ionization threshold, $\varepsilon=\sqrt{p^2c^2+c^4}-c^2$. The orbital angular
momentum of the partial wave is denoted by $l$ and the corresponding magnetic
quantum number $m_l$ and the phases $\Delta_{\kappa}$ are chosen so that the
continuum wave function fulfills the boundary conditions of an incoming plane
wave and an outgoing spherical wave.  The total angular momentum quantum number
of the partial wave is $j=|\kappa|-\frac{1}{2} $. The $Y_{lm_l}(\Omega_e)$
denote the spherical harmonics and the $C\left(l\ \frac{1}{2}\ j;m_l\ m_s\
m\right)$ stand for the vector coupling coefficients.

For describing the nucleus we have used a collective model \cite{Greiner} in
which the excitations of the nucleus are assumed to be vibrations of the nuclear
surface. The surface can be parametrized as
\begin{equation}
R(\theta,\varphi,t)=R_0\Big(1+\sum_{\lambda=0}^\infty\sum_{\mu=-\lambda}^{\lambda} \alpha_{\lambda\mu}^*(t)Y_{\lambda\mu}(\Omega)\Big)\ .
\end{equation}
$R_0$ denotes the radius of a homogeneously charged sphere and the time-dependent amplitudes $\alpha_{\lambda\mu}$ serve as collective coordinates.
Using this parametrization and requiring that the charge is homogeneously
distributed, the nuclear charge density can be written as
\begin{equation}
\rho_n(\vec{r},t)=\rho_0\theta\left(R(\theta,\varphi,t)-r\right)\ ,
\end{equation}
with the constant average density $\rho_0=\frac{3Ze}{4\pi R_0^3}$. In the first
order in the collective coordinates the density $\rho_n$ can be approximated as
\begin{equation}
\rho_n (\vec{r},t)=\rho_0 \theta(R_0-r)+\rho_0R_0\delta(R_0-r)\sum_{\lambda\mu}\alpha_{\lambda\mu}^*(t)Y_{\lambda\mu}(\Omega)\ .
\label{rho}
\end{equation}
The electron-nucleus interaction Hamiltonian in eq.\ (\ref{eq:coulomb}) can be
written using the multipole expansion as
\begin{equation}
H_{en}=\sum_{L=0}^\infty\sum_{M=-L}^L\frac{4\pi}{2L+1}Y^*_{LM}(\Omega_e)
\int d^3r\frac{r^{L}_<}{r^{L+1}_>}Y_{LM}(\Omega_n)\rho_n(\vec{r}_n) \ .
\end{equation}
The radius $r_e$($r_n$) denotes the electronic (nuclear) radial coordinate and
$\Omega_e$($\Omega_n$) stands for the corresponding solid angle. We can make the
simplifying assumption that the electron does not enter the nucleus, therefore we
assume that $r_e>r_n$. According to Ref. \cite{Rose}, this approximation should
not affect the results for the considered transitions. The Hamiltonian can then be written
\begin{equation}
H_{en}=\sum_{L=0}^\infty\sum_{M=-L}^L\frac{4\pi}{2L+1}Y^*_{LM}(\Omega_e)\frac{1}{r_e^{L+1}}Q_{LM}\ ,
\end{equation}
with the help of the electric multipole moments
\begin{equation}
Q_{LM}=\int d^3r_n r_n^{L} Y_{LM}(\Omega_n)\rho_n(\vec{r}_n)\ .
\end{equation}

After performing some angular momentum algebra one finds for the matrix element
of $H_{en}$ for a given partial wave component
\begin{eqnarray}
&&\langle N_d^*I_dM_{I_d}|\langle \kappa_dm_d|H_{en}|\kappa m\rangle|N_iI_i M_{I_i}\rangle
=
\nonumber \\
&&\sum_{LM}(-1)^{I_d+M_{I_i}+L+M+m+3j_d}R^{(1)}_{L,j_d,j} 
\langle N^* I_d\|Q_L\|NI_i\rangle\sqrt{2j_d+1}\sqrt{\frac{4\pi}{(2L+1)^3}}
 \nonumber \\ 
&&\times
C(I_i\ I_d\ L;-M_{I_i}\ M_{I_d} \ M)
 \ C(j\ j_d\ L;-m\ m_d\ M)\
C\left(j_d\ L\ j;\frac{1}{2}\ 0 \ \frac{1}{2}\right)\, .
\end{eqnarray}
$R^{(1)}_{L,j_d,j}$ stands for the radial integral given by
\begin{equation}
R^{(1)}_{L,j_d,j}=\int_0^\infty drr^{-L+1}\Big(f_{\kappa_d}(r)f_{\varepsilon\kappa}(r)+g_{\kappa_d}(r)g_{\varepsilon\kappa}(r)\Big)
\ ,
\label{Radial}
\end{equation}
where $g(r)$ and $f(r)$ are the large and small radial components of the
relativistic continuum electron wave function
\begin{equation}
\Psi_{\varepsilon\kappa m}(\vec{r})=\left(\begin{array}{c} g_{\varepsilon\kappa}(r)\Omega_{\kappa}^{m}(\Omega_e)\\if_{\varepsilon\kappa}(r)\Omega_{-\kappa}^{m}(\Omega_e)\end{array}
\right)\ ,
\end{equation}
and the $g_d(r)$ and $f_d(r)$ are the components of the bound Dirac wave functions
\begin{equation}
\Psi_{n_d\kappa_dm_d}(\vec{r})=\left(\begin{array}{c} g_{\kappa_d}(r)\Omega_{\kappa_d}^{m_d}(\Omega_e)\\if_{\kappa_d}(r)\Omega_{-\kappa_d}^{m_d}(\Omega_e)\end{array}
\right)\ 
\end{equation}
with the spherical spinor functions $\Omega_{\kappa}^{m}$.
For a given  multipolarity  $L$, the NEEC rate for an electric transition thus reads
\begin{equation}
\label{aprate}
Y^{(e)}_n=\frac{4\pi^2\rho_i}{(2L+1)^2} B\uparrow (EL,I_i\to I_d)(2j_d+1)
\sum_\kappa |R^{(1)}_{L,j_d,j}|^2\
C\left(j_d\ L\ j;\frac{1}{2}\ 0\ \frac{1}{2}\right)^2\ , 
\end{equation}
where 
\begin{equation}
B\uparrow (EL,I_i\to  I_d)=\frac{1}{2I_i+1}|\langle N^* I_d\|Q_L\|NI_i\rangle |^2
\end{equation}
is the reduced electric transition probability.

If we consider the charge density of the nuclear collective model from eq.\ (\ref{rho}),
the matrix element of the Hamiltonian $H_{en}$ can be conveniently written in
terms of the reduced transition probability $B\uparrow$ without imposing any
constraints on the electron motion. In this case the electric NEEC rate is given by
\begin{eqnarray}
\label{rate}
Y^{(e)}_n&=&\frac{4\pi^2\rho_i}{(2L+1)^2}R_0^{-2(L+1)} B\uparrow (EL,I_i\to I_d)(2j_d+1)
\nonumber \\ &\times&
\sum_\kappa |R^{(2)}_{L,j_d,j}|^2\
C\left(j_d\ L\ j;\frac{1}{2}\ 0\ \frac{1}{2}\right)^2\ , 
\end{eqnarray}
where the electronic radial integral is
\begin{eqnarray}
R^{(2)}_{L,j_d,j}&=&\frac{1}{R_0^{L-1}}\int_0^{R_0} dr r^{L+2}\left(f_{\kappa_d}(r)f_{\varepsilon\kappa}(r)+
g_{\kappa_d}(r)g_{\varepsilon\kappa}(r)\right)+ \nonumber \\
&+&R_0^{L+2}\int_{R_0}^\infty dr r^{-L+1}\left(f_{\kappa_d}(r)f_{\varepsilon\kappa}(r)+g_{\kappa_d}(r)g(r)_{\varepsilon\kappa}\right)\ .
\label{rrs}
\end{eqnarray}

The magnetic Hamiltonian in eq.\ (\ref{hmagn}) can be written using the multipole
expansion as
\begin{equation}
H_{magn}=-\vec{\alpha}\cdot\vec{A}=-\frac{1}{c}\sum_{LM}\frac{4\pi}{2L+1}\vec{\alpha}\cdot
\vec{Y}^M_{LL}(\Omega_e) \int
d^3r_n\frac{r_<^L}{r_>^{L+1}}\vec{j}_n(\vec{r}_n)\cdot\vec{Y}^{M*}_{LL}(\Omega_n)\ .
\end{equation}
We use again the approximation that the electron does not enter the nucleus.
Then the Hamiltonian reads
\begin{equation}
 H_{magn}=-i\sum_{LM}\frac{4\pi}{2L+1}\sqrt{\frac{L+1}{L}}
r_e^{-(L+1)}M_{LM}\vec{\alpha}\cdot\vec{Y}^{M*}_{LL}(\Omega_e)\ ,
\label{hmagnmult}
\end{equation}
where  the magnetic multipole operator is given by \cite{Schwartz} 
\begin{equation}
M_{LM}=-\frac{i}{c}\sqrt{\frac{L}{L+1}}\int d^3r
r^L\vec{Y}^{M}_{LL}(\Omega_n)\cdot\vec{j}_n(\vec{r}_n)\ .
\end{equation}
Here, the vector spherical harmonics are defined as \cite{Edmonds}
\begin{equation}
\vec{Y}^M_{LL}(\Omega_e)=\sum_\nu\sum_q C(L\ 1\ L;\nu\ q\ M)Y_{L\nu}(\Omega_e)\vec{\epsilon}_q\ ,
\label{vsh}
\end{equation}
where $q=0,\pm 1$ and the spherical vectors $\vec{\epsilon}_q$ are
\begin{eqnarray}
&\vec{\epsilon}_+&=-\frac{1}{\sqrt{2}}(\vec{e}_x+i\vec{e}_y)\ ,\nonumber\\
&\vec{\epsilon}_0&=\vec{e}_z \ ,\nonumber\\ 
&\vec{\epsilon}_-&=\frac{1}{\sqrt{2}}(\vec{e}_x-i\vec{e}_y)\ .
\end{eqnarray}
Introducing the expression for the vector spherical harmonics in eq.\
(\ref{hmagnmult}) we obtain
\begin{eqnarray} 
H_{magn}&=&i\sum_{LM}\frac{4\pi(-1)^M}{2L+1}\sqrt{\frac{L+1}{L}}
r^{-(L+1)}M_{LM}\nonumber \\
&\times&\sum_\nu Y_{L\nu}(\Omega_e)
\Bigg(-\frac{1}{\sqrt{2}}C(L\ 1\ L;\nu\ 1\ -M)(\alpha_x+i\alpha_y)\nonumber \\
&+&C(L\ 1\ L;\nu\ 0\ -M)\alpha_z+
\frac{1}{\sqrt{2}}C(L\ 1\ L;\nu\ -1\ -M)(\alpha_x-i\alpha_y)\Bigg)\ ,
\end{eqnarray}
where $\alpha_x$, $\alpha_y$ and $\alpha_z$ are the Cartesian $\vec{\alpha}$ matrices. The matrix element of the magnetic Hamiltonian  then yields 
\begin{eqnarray}
&&\langle N^*I_dM_{I_d}|\langle n_d\kappa_dm_d|-\vec{\alpha}\cdot\vec{A}|\varepsilon\kappa m\rangle|N I_i M_{I_i}\rangle=\nonumber \\
&&4\pi i\sum_{LM\nu}(-1)^M\sqrt{\frac{L+1}{L}}\frac{1}{2L+1}
\langle N^* I_d M_{I_d}|M_{LM}|N I_i M_{I_i}\rangle 
\nonumber \\
&&\times
\Bigg(-\frac{1}{\sqrt{2}}C(L\ 1\ L;\nu\ 1\ -M)\langle
n_d\kappa_dm_d|r^{-(L+1)}Y_{L\nu}(\Omega_e)(\alpha_x+i\alpha_y)|\varepsilon\kappa
m\rangle \
\nonumber \\
&&+ C(L\ 1\ L;\nu\ 0\ -M)\langle
n_d\kappa_dm_d|r^{-(L+1)}Y_{L\nu}(\Omega_e)\alpha_z|\varepsilon\kappa m\rangle \
\nonumber \\
&&+\frac{1}{\sqrt{2}}C(L\ 1\ L;\nu\ -1\ -M)\langle n_d\kappa_dm_d|r^{-(L+1)}Y_{L\nu}(\Omega_e)(\alpha_x-i\alpha_y)|\varepsilon\kappa m\rangle \Bigg)\ .
\label{aA1}
\end{eqnarray}

Introducing the notations $T^+_{di,\nu}$, $T^0_{di,\nu}$ and $T^-_{di,\nu}$ for the electronic matrix
elements we get
\begin{eqnarray}
&&\langle N^*I_dM_{I_d}|\langle n_d\kappa_dm_d|-\vec{\alpha}\cdot\vec{A}|\varepsilon\kappa m\rangle|N I_i M_{I_i}\rangle=\nonumber \\ 
&&4\pi i
\sum_{LM\nu}(-1)^{I_i-M_{I_i}+M}
\sqrt{\frac{L+1}{L(2L+1)^3}}\ C(I_d\ I_i\ L;M_{I_d}\ -M_{I_i}\ M)\langle N^* I_d ||M_L||N I_i\rangle\nonumber \\
&&\times
\Bigg(-\frac{1}{\sqrt{2}}\ C(L\ 1\ L;\nu\ 1\ -M)\ T^+_{di,\nu}
+C(L\ 1\ L;\nu\ 0\ -M)\ T^0_{di,\nu} \nonumber \\
&&+\frac{1}{\sqrt{2}}\ C(L\ 1\ L;\nu \ -1\ -M)\ T^-_{di,\nu}\Bigg)\ .
\end{eqnarray}
The explicit form of the electronic matrix elements can be found in Appendix
\ref{B}. For a given multipolarity $L$, the rate for the nuclear excitation by
electron capture for a pure magnetic transition is then
\begin{eqnarray}
Y_n^{(m)}&=&\frac{16\pi^3(L+1)\rho_i}{L(2L+1)^3}B\uparrow(ML,I_i\to I_d)\sum_{\kappa m m_d}
\nonumber \\ 
&\times&
\Big|\sum_{M\nu}\Big(-\frac{1}{\sqrt{2}}\ C(L\ 1\ L;\nu\ 1\ -M)\ T^+_{di,\nu}
+C(L\ 1\ L;\nu\ 0\ -M)\ T^0_{di,\nu} \nonumber \\
&+&\frac{1}{\sqrt{2}}\ C(L\ 1\ L;\nu\ -1\ -M)\ T^-_{di,\nu}\Big)\Big|^2\ .
\end{eqnarray}
All the nuclear information is contained in the reduced magnetic transition
probability
\begin{equation}
B\uparrow(ML,I_i\to I_d)=\frac{1}{2I_i+1}|\langle N^* I_d\|M_L(t)\|N I_i\rangle|^2\ ,
\end{equation} 
whose value can be taken from experimental data or from calculations involving
different nuclear models. Given the different parity of the electric and
magnetic multipole moments a transition of a given multipolarity $L$ is either
electric or magnetic. We consider only the cases of transitions with a certain
value of $L$ and we neglect the possible mixing ratios between electric and
magnetic multipoles of different multipolarities.


\section{\label{results} Numerical results}

We calculate total cross sections and resonance strengths for NEEC followed by
the radiative decay of the excited nucleus for various collision systems,
involving electric $E2$ and magnetic $M1$ multipole transitions.  We consider the
cases of nuclear isotopes with low-lying nuclear levels for which the NEEC
process is more likely to be observed experimentally.

For the case of electric multipole transitions we have considered the $0^+\to
2^+$ $E2$ transitions of the $^{236}_{92}\mathrm{U}$, $^{238}_{92}\mathrm{U}$,
$^{248}_{96}\mathrm{Cm}$, $^{174}_{70}\mathrm{Yb}$, $^{170}_{68}\mathrm{Er}$,
$^{154}_{64}\mathrm{Gd}$, $^{156}_{64}\mathrm{Gd}$, $^{162}_{66}\mathrm{Dy}$ and
$^{164}_{66}\mathrm{Dy}$ even-even nuclei.  The quadrupole excitations of
even-even nuclei are well described by the collective model. For the calculation
of the NEEC rate both formulas from eq.\ (\ref{aprate}) and eq.\ (\ref{rate}) have
been used for a comparison.  A further $E2$ transition
$\frac{5}{2}^-\to\frac{7}{2}^-$ of the odd $^{163}_{66}\mathrm{Dy}$ nucleus has
been investigated. For this case we have calculated the NEEC rate assuming that
the electron does not enter the nucleus. The reduced transition probability
$B\uparrow(E2)$ for the even-even nuclei as well as the energies of the nuclear
levels were taken from \cite{Raman}, and in the case of the $^{163}_{66}\mathrm{Dy}$ isotope, from \cite{NDS10}. The nuclear radiative rate was calculated
according to the formula \cite{Ring}
\begin{equation}
A_r^{d \to f}(\lambda,L)=\frac{8\pi(L+1)}{L((2L+1)!!)^2}\frac{E^{2L+1}}{c}B\downarrow(\lambda L,I_d\to
I_f)\ ,
\end{equation}
where $\lambda=E,M$ stands for the type of transition. The two reduced transition
probabilities for the emission, respectively the absorption of a gamma ray are
related through the formula
\begin{equation}
B\downarrow(\lambda L,I_d\to I_f)=\frac{2I_f+1}{2I_d+1}B\uparrow(\lambda L,I_f\to I_d)\ .
\end{equation}
The width of the excited nuclear state is then
\begin{equation}
\label{width}
\Gamma_d=A_r^{d \to f}+A^d_{\rm IC} \,
\end{equation}
where $A^d_{\rm IC}$ is the IC rate of the state $d$, related to the NEEC rate
through the principle of detailed balance,
\begin{equation}
A^d_{\rm IC}=\frac{2(2I_i+1)}{(2I_d+1)(2j_d+1)}Y_n \ .
\end{equation}

For the NEEC rate we need to calculate numerically the radial integrals
$R_{L,j_d,j}$  that enter eq.\ (\ref{aprate}) and eq.\ (\ref{rate}). We use
relativistic Coulomb-Dirac wave functions for the continuum electron and wave
functions calculated with the GRASP92 package \cite{Par96} considering a
homogeneously charged nucleus for the bound electron. The value of $R_{L,j_d,j}$
is almost the same whether we use Coulomb-Dirac radial wave functions or we take
into account the finite size of the nucleus. The finite size of the
nucleus has a sensitive effect on the energy levels of the bound
electron. The energy level of the bound electron is calculated with GRASP92 and
it includes quantum electrodynamic corrections. The first term of the 
sum of radial integrals in eq.\
(\ref{rrs}) is about 3 orders of magnitude smaller than the second term,
\begin{eqnarray}
&& \frac{1}{R_0^{L-1}}\int_0^{R_0} dr r^{L+2}\left(f_{\kappa_d}(r)f_{\varepsilon\kappa}(r)+g_{\kappa_d}(r)g_{\varepsilon\kappa}(r)\right)
 \nonumber \\
&&\ll
R_0^{L+2}\int_{R_0}^\infty dr r^{-L+1}\left(f_{\kappa_d}(r)f_{\varepsilon\kappa}(r)+g_{\kappa_d}(r)g_{\varepsilon\kappa}(r)\right)\ .
\end{eqnarray}
Here, the nuclear radius $R_0$ is calculated according to the formula \cite{Soff}
\begin{equation}
R_0=(1.0793A^{1/3}+0.73587)\mathrm{fm}\ ,
\end{equation}
where $A$ is the atomic mass number. If we make the approximation
\begin{equation}
R^{(2)}_{L,j_d,j}\simeq R_0^{L+2}\int_0^\infty dr r^{-L+1}\left(f_{\kappa_d}(r)f_{\varepsilon\kappa}(r)+g_{\kappa_d}(r)g_{\varepsilon\kappa}(r)\right) \,
\end{equation}
the NEEC rate is exactly the one in (\ref{aprate}), calculated with the
assumption that the electron does not enter the nucleus. For the numerical cases
of the even-even nuclei the difference between the rates calculated with eq.\
(\ref{aprate}) and eq.\ (\ref{rate}) are from under 1$\%$ up to 6$\%$. The
difference is larger for the capture into the $s$ orbitals and it increases with
the value of the atomic number Z. For the capture of the continuum electron into
the $2s$ orbital of $^{248}_{96}\mathrm{Cm}$, the value of the rate calculated using the non-penetrating approximation
 is 6$\%$ larger than the one
calculated with the collective model, in which the restriction on the electron motion is avoided.

 \begin{table*}
\caption{\label{Etable} Electric NEEC rates and resonance strengths for various heavy ion
collision systems. $E_\mathrm{exc}$ is the nuclear excitation energy,
$E_{\mathrm{c}}$ is the continuum electron energy and $\Gamma_{\mathrm{d}}$ is the width of
the excited nuclear state.}
\begin{ruledtabular}
\begin{tabular}{lrrccccc}

Isotope & $E_{\mathrm{exc}}$(keV) & $E_{\mathrm{c}}$(keV) &Type & Orbital &$Y_{\mathrm{n}}(1/s)$&$\Gamma_{\mathrm{d}}$(eV)& $S$(barn$\cdot$eV) \\ 
\hline

$^{164}_{66}\mathrm{Dy}$ & 73.392 &10.318& E2 & $1s_{1/2}$ &$1.86\cdot10^8$ &$4.36\cdot 10^{-8}$   &$3.88\cdot10^{-2}$\\

$^{170}_{68}\mathrm{Er}$ & 78.591 &11.350& E2 & $1s_{1/2}$ &$2.22\cdot10^8$ & $5.74\cdot 10^{-8}$  &$4.70\cdot10^{-2}$\\

$^{174}_{70}\mathrm{Yb}$ & 76.471 &4.897& E2 & $1s_{1/2}$ &$1.78\cdot10^8$ &$4.84\cdot 10^{-8}$   &$9.25\cdot10^{-2}$\\

$^{154}_{64}\mathrm{Gd}$ & 123.071 &64.005& E2 & $1s_{1/2}$ &$5.67\cdot10^8$ &$2.51\cdot 10^{-7}$   &$2.90\cdot10^{-2}$\\

$^{156}_{64}\mathrm{Gd}$ & 88.966 &74.742& E2 & $2s_{1/2}$ &$3.34\cdot10^7$ & $1.21\cdot 10^{-7}$  &$7.07\cdot10^{-4}$ \\
$^{156}_{64}\mathrm{Gd}$ & 88.966&74.896& E2 & $2p_{1/2}$ & $1.17\cdot10^8$& $1.32\cdot 10^{-7}$  &$2.26\cdot10^{-3}$\\
$^{156}_{64}\mathrm{Gd}$ & 88.966&75.680& E2 & $2p_{3/2}$ & $1.60\cdot10^8$& $1.27\cdot 10^{-7}$ &$3.17\cdot10^{-3}$\\

$^{162}_{66}\mathrm{Dy}$ & 80.660 &65.432& E2 & $2s_{1/2}$ &$2.80\cdot10^7$ & $9.39\cdot 10^{-8}$  &$6.23\cdot10^{-4}$ \\
$^{162}_{66}\mathrm{Dy}$ & 80.660 &66.594& E2 & $2p_{1/2}$ & $1.60\cdot10^8$& $1.11\cdot 10^{-7}$  &$2.99\cdot10^{-3}$\\
$^{162}_{66}\mathrm{Dy}$ & 80.660 &66.492& E2 & $2p_{3/2}$ & $2.16\cdot10^8$& $1.04\cdot 10^{-7}$ &$4.25\cdot10^{-2}$\\

$^{163}_{66}\mathrm{Dy}$ & 73.440 &58.212& E2 & $2s_{1/2}$ &$9.18\cdot10^6$ & $1.66\cdot 10^{-7}$  &$1.33\cdot10^{-4}$ \\
$^{163}_{66}\mathrm{Dy}$ & 73.440 &58.374& E2 & $2p_{1/2}$ & $6.93\cdot10^7$& $1.96\cdot 10^{-7}$  &$8.54\cdot10^{-4}$\\
$^{163}_{66}\mathrm{Dy}$ & 73.440 &58.272& E2 & $2p_{3/2}$ & $9.44\cdot10^7$& $1.85\cdot 10^{-7}$ &$1.21\cdot10^{-3}$\\

$^{236}_{92}\mathrm{U}$ & 45.242 &11.113& E2 & $2s_{1/2}$ &$1.16\cdot10^8$ & $1.89\cdot 10^{-8}$  &$8.79\cdot10^{-3}$ \\
$^{236}_{92}\mathrm{U}$ & 45.242 &11.038& E2 & $2p_{1/2}$ & $3.16\cdot10^{9}$& $4.19\cdot 10^{-7}$  &$1.05\cdot10^{-2}$\\
$^{236}_{92}\mathrm{U}$ & 45.242 &15.601& E2 & $2p_{3/2}$ & $3.22\cdot10^{9}$& $2.16\cdot 10^{-7}$ &$1.56\cdot10^{-2}$\\

$^{238}_{92}\mathrm{U}$ & 44.910 & 10.782 &E2  & $2s_{1/2}$ &$1.17\cdot10^8$ &$1.90\cdot 10^{-8}$
& $8.90\cdot10^{-3}$\\
$^{238}_{92}\mathrm{U}$ & 44.910 &10.706 &E2 & $2p_{1/2}$ &$3.20\cdot10^9$ &$4.25\cdot 10^{-7}$
&$1.06\cdot10^{-2}$\\
$^{238}_{92}\mathrm{U}$ & 44.910 &15.269 &E2& $2p_{3/2}$ & $3.27\cdot10^9$ &$2.19\cdot 10^{-7}$
  &$1.56\cdot10^{-2}$ \\
 
$^{248}_{96}\mathrm{Cm}$ & 43.380 &5.500& E2 & $2s_{1/2}$ &$2.32\cdot10^8$  &$3.42\cdot 10^{-8}$
 &$1.79\cdot10^{-2}$\\
$^{248}_{96}\mathrm{Cm}$ & 43.380 &5.398& E2 & $2p_{1/2}$ & $5.61\cdot10^9$&$7.42\cdot 10^{-7}$
  &$1.91\cdot10^{-2}$\\
$^{248}_{96}\mathrm{Cm}$ & 43.380 & 11.018&E2 & $2p_{3/2}$ & $5.42\cdot10^9$&$3.60\cdot 10^{-7}$
  &$2.20\cdot10^{-2}$\\
\end{tabular}
\end{ruledtabular}
\end{table*}

For the cases of the $\mathrm{U}$ isotopes and for $^{248}_{96}\mathrm{Cm}$, the
capture into the K shell is not possible due to the low energy level of the first
excited nuclear state. For these 3 isotopes, recombination into the L shell of
initially He-like ions is the most probable one. We regard the capture of the electron into a closed shell configuration as a one-electron problem, without the participation of the K-shell electrons.
We consider for the continuum electron a total screening, while the bound electron is described by wave functions for an extended nucleus. 
The electron interaction is included in the bound radial wave functions calculated with GRASP92 and it 
influences the results through the value of the bound energy and through the
shape of the electronic radial wave functions. The change of the shape of the radial
wave functions has a larger numerical contribution to the value of the NEEC
rate than the change of the energy due to the electron-electron interaction.

For the other cases of even-even nuclei, capture into the K shell
is possible. For the $^{156}_{64}\mathrm{Gd}$, $^{162}_{66}\mathrm{Dy}$, and
$^{163}_{66}\mathrm{Dy}$ isotopes we have also considered the capture into the He-like ions. The width of the nuclear state in eq.\ (\ref{width}) contains then an extra term which accounts for the possible IC of the K-shell electrons. 
The capture rate into the $2p$
orbitals is in general one order of magnitude larger than the one for the capture into the
$2s$ orbital. 
The NEEC rates and resonance strengths for NEEC followed by the radiative decay
of the nucleus for electric transitions are presented in Table \ref{Etable} .
The values of the NEEC rates have been calculated using the non-penetrating
approximation.

\begin{table*}
\caption{\label{Mtable} Magnetic NEEC rates and resonance strengths for various heavy ion
collision systems. $E_\mathrm{exc}$ is the nuclear excitation energy,
$E_{\mathrm{c}}$ is the continuum electron energy and $\Gamma_{\mathrm{d}}$ is the width of
the excited nuclear state.}
\begin{ruledtabular}
\begin{tabular}{lrrccccc}

Isotope & $E_{\mathrm{exc}}$(keV) & $E_{\mathrm{c}}$(keV) &Type & Orbital &$Y_{\mathrm{n}}(1/s)$&$\Gamma_{\mathrm{d}}$(eV)& $S$(barn$\cdot$eV) \\ 
\hline

$^{165}_{67}\mathrm{Ho}$ &94.700 & 29.563& M1 & $1s_{1/2}$ & $4.50\cdot 10^9$& $7.33\cdot 10^{-6}$ & $4.95\cdot10^{-1}$\\

$^{173}_{70}\mathrm{Yb}$ & 78.647 & 7.073& M1 & $1s_{1/2}$ &  $2.52\cdot 10^9$&$2.43 \cdot 10^{-6}$  &$8.57\cdot10^{-1}$
\\

$^{185}_{75}\mathrm{Re}$ & 125.358 & 42.198& M1 & $1s_{1/2}$ &  $9.20\cdot 10^9$&$1.51\cdot 10^{-5}$  &$7.33\cdot10^{-1}$\\

$^{187}_{75}\mathrm{Re}$ & 134.243 &51.083 & M1 & $1s_{1/2}$ &  $8.82\cdot 10^9$& $1.67\cdot 10^{-5}$ & $6.05\cdot10^{-1}$\\

$\ ^{55}_{25}\mathrm{Mn}$ &125.949 & 117.378& M1 & $1s_{1/2}$ & $1.33\cdot 10^7$& $1.75\cdot 10^{-6}$ & $5.02\cdot10^{-4}$\\

$^{57}_{26}\mathrm{Fe}$ & 14.412 & 5.135& M1 & $1s_{1/2}$ &  $2.24\cdot 10^6$&$1.25 \cdot 10^{-9}$  &$8.83\cdot10^{-4}$
\\

$^{40}_{19}\mathrm{K}$ & 29.829 & 24.896& M1 & $1s_{1/2}$ &  $6.03\cdot 10^6$&$8.85\cdot 10^{-8}$  &$1.10\cdot10^{-3}$\\
 
$^{155}_{64}\mathrm{Gd}$ & 60.008 &45.784 & M1 & $2s_{1/2}$ &  $9.38\cdot 10^8$& $8.21\cdot 10^{-7}$ & $2.62\cdot10^{-3}$\\
$^{155}_{64}\mathrm{Gd}$ & 60.008 &45.938& M1 & $2p_{1/2}$ &  $1.04\cdot 10^7$& $7.84\cdot 10^{-7}$ & $3.05\cdot10^{-4}$\\
$^{155}_{64}\mathrm{Gd}$ & 60.008 &46.722& M1 & $2p_{3/2}$ &  $3.75\cdot 10^6$& $7.80\cdot 10^{-7}$ & $1.08\cdot10^{-4}$\\

$^{157}_{64}\mathrm{Gd}$ & 54.533 &40.309 & M1 & $2s_{1/2}$ &  $1.42\cdot 10^8$& $3.17\cdot 10^{-7}$ & $1.34\cdot10^{-2}$\\
$^{157}_{64}\mathrm{Gd}$ & 54.533 &40.463 & M1 & $2p_{1/2}$ &  $1.59\cdot 10^7$& $2.61\cdot 10^{-7}$ & $1.82\cdot10^{-3}$\\

$^{157}_{64}\mathrm{Gd}$ & 54.533 &41.247 & M1 & $2p_{3/2}$ &  $5.81\cdot 10^6$& $2.56\cdot 10^{-7}$ & $6.67\cdot10^{-4}$
\end{tabular}
\end{ruledtabular}
\end{table*}

For the magnetic multipole transitions we consider the $M1$ transitions of the
odd isotopes $\ ^{165}_{67}\mathrm{Ho}$, $^{173}_{70}\mathrm{Yb}$,
$^{55}_{25}\mathrm{Mn}$, $^{57}_{26}\mathrm{Fe}$, $^{40}_{19}\mathrm{K}$,
$^{155}_{64}\mathrm{Gd}$, $^{157}_{64}\mathrm{Gd}$, $^{185}_{75}\mathrm{Re}$ and
$^{187}_{75}\mathrm{Re}$. Numerical results for these ions are presented
in Table \ref{Mtable}. The electronic radial integrals are calculated
numerically using the same type of wave functions for the bound and continuum
electron as for the electric transitions. The reduced magnetic transition
probability $B\downarrow(M1)$ and the energies of the nuclear levels are taken
from \cite{NDS1,NDS2,NDS3,NDS4,NDS5,NDS6,NDS7,NDS8,NDS9}. Capture into the K shell is possible
for all the chosen ions, except for the $^{157}_{64}\mathrm{Gd}$ isotope. Capture into the higher shells is less probable, and
already for the capture into the $2s$ orbital of $\ ^{167}_{67}\mathrm{Ho}$, the
NEEC rate is one order of magnitude smaller. We present also results for captures into the He-like ions of $^{155}_{64}\mathrm{Gd}$ and $^{157}_{64}\mathrm{Gd}$.
The largest resonance strength is the one for the capture into the $1s$
orbital of $^{173}_{70}\mathrm{Yb}$, namely, $S=8.57\cdot10^{-1}$ barn$\cdot$eV.
This value is small in comparison with the DR resonance strengths, which are in
the order of magnitude of $10^3$ barn$\cdot$ eV.

\begin{figure}
\includegraphics{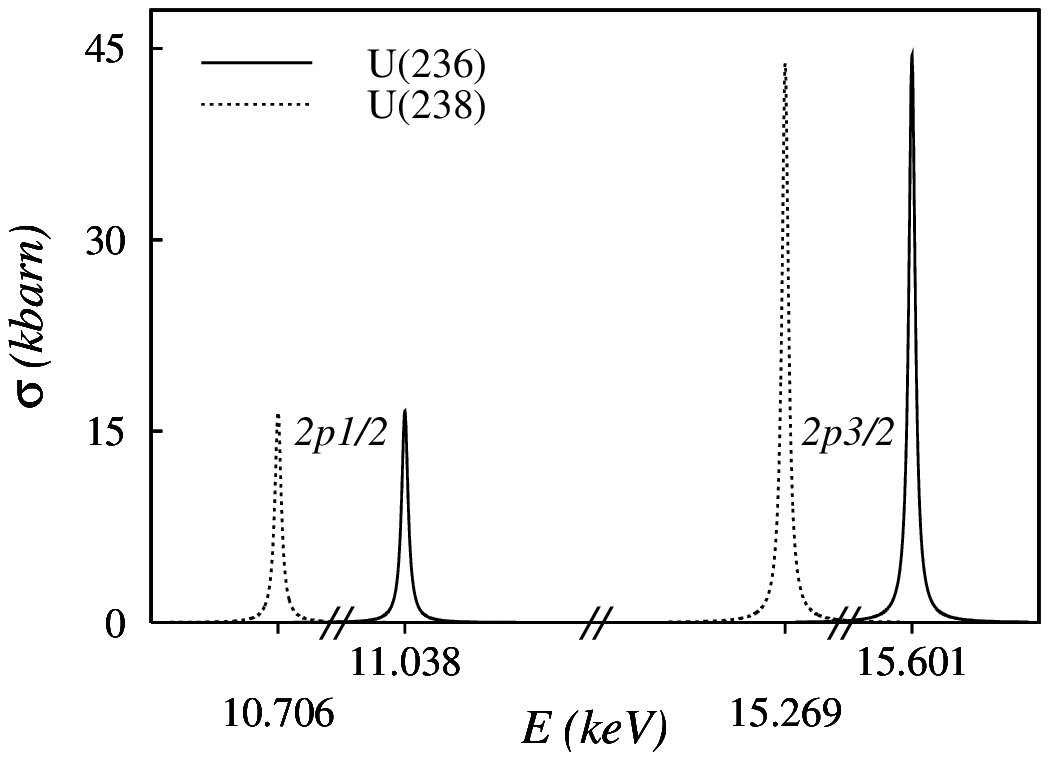}
\caption{\label{Uplot}NEEC cross sections for Uranium isotopes as a function of the continuum electron energy}
\end{figure}

In Figure \ref{Uplot} we present the cross section for the capture of the
continuum electron into the $2p_{1/2}$ and $2p_{3/2}$ orbitals of the two
studied even-even uranium isotopes. The cross sections for the capture into the
$2p_{3/2}$ orbitals are larger than the ones into the $2p_{1/2}$
orbitals. Although the cross section values are in the order of thousands of
barns, the width of the Lorentzian is given by the width of the excited nuclear
state, which is in the order of $10^{-7}$ eV. This validates the use of the 
isolated resonance approximation in Section \ref{pertexp}.

\section{\label{sum} Summary}

In this article we present a versatile formalism for describing complex
processes actively involving atomic electrons and nuclei and derive a total
cross section formula for the process of NEEC. The cross section can be traced
back to the calculation of IC rates and radiative transition
rates. It also requires the knowledge of nuclear excitation energies and the
bound state energy of the electron after capture.

We derive NEEC rates for the case of both electric and magnetic multipole transitions
using relativistic electronic wave functions. For the bound electron we
use radial wave functions which take into account the finite size of the
nucleus. The nucleus is described using a nuclear collective model. The
nuclear part of the matrix element is written with the help of the reduced
nuclear transition probability whose value is taken from experiment. 

We calculate numerically the cross sections for NEEC followed by the radiative
decay of the excited nuclear state for various nuclei. Particular interest is
payed to the cases that are more likely to be observed experimentally. The
obtained resonance strengths are typically 4 orders of magnitude smaller
than the DR ones, which is due to the small width of the excited nuclear states. 

If the initial and final states for NEEC are the same as the ones for the process of radiative recombination (RR), quantum interference
between the two processes occurs. The cross section for RR is much larger than the
one of NEEC, therefore, the inclusion of interference terms is expected to increase the 
cross section by several orders of magnitude, making the experimental observation
of NEEC feasible. Calculations about the magnitude of this effect are in progress.


\begin{acknowledgments}

The authors would like to thank Prof. R. V. Jolos for enlightening discussions
regarding nuclear structure theory. Z. H. also appreciates useful discussions with
Prof. H. Tawara and J. R. Crespo L\'opez-Urrutia about experimental issues of
recombination processes. A. P.  acknowledges the financial
support from the Deutsche Forschungsgemeinschaft (DFG).

\end{acknowledgments}

\appendix
\section{\label{A}}

In this Appendix we show that the term $QH_{nr}RG_0RH_{er}P$ can be approximated
by $QH_{magn}P$, where we introduced the Hamiltonian
\begin{equation}\label{eq:hmagn}
H_{magn} = - \frac{1}{c} \vec{\alpha} \int d^3r_n \frac{\vec{j}_n(\vec{r}_n)}{|\vec{r}-\vec{r}_n|}\ ,
\end{equation}
describing the magnetic interaction of the nuclear and electric currents due to
the exchange of a transverse photon. The replacement is valid in the case when
the frequency of the virtual photon is negligible, or, in other terms, when its
wavelength is large compared to the typical linear size of the total system.

In this derivation we use the second quantized forms
\begin{eqnarray}\label{eq:her}
H_{er} &=& \sum_{ab}\sum_{\sigma} \int d^3k \frac{1}{2\pi} \sqrt{\frac{c}{k}} c_a^{\dagger} c_b
\nonumber  \\
&\times&\int d^3 r_e \phi_a^*(\vec{r}_e) \vec{\alpha} 
\left( 
\vec{\epsilon}_{\vec{k}\sigma} a_{\vec{k}\sigma} e^{i\vec{k}\cdot\vec{r}_e} + 
\vec{\epsilon}_{\vec{k}\sigma}^* a_{\vec{k}\sigma}^{\dagger} e^{-i\vec{k}\cdot\vec{r}_e}
\right) \phi_b(\vec{r}_e) 
\end{eqnarray}
and
\begin{eqnarray}\label{eq:hnr}
H_{nr} &=& - \frac{1}{(2\pi)^{5/2}} \sqrt{\frac{1}{ck}} \sum_{st}\sum_{\sigma'} \int d^3k' B_s^{\dagger} B_t
\nonumber \\ 
&\times&
\int d^3 r_n \vec{j}_n 
\left( 
\vec{\epsilon}_{\vec{k}'\sigma'} a_{\vec{k}'\sigma'} e^{i\vec{k}'\cdot\vec{r}_n} + 
\vec{\epsilon}_{\vec{k}'\sigma'}^* a_{\vec{k}'\sigma'}^{\dagger} e^{-i\vec{k}'\cdot\vec{r}_n}
\right)
\end{eqnarray}
of the electromagnetic interaction Hamiltonians. The $\phi_a$ form a complete
set of one-electron states, and the $c_a^{\dagger}$ ($c_b$) are electronic creation
(annihilation) operators. The $\psi_s$ and $\psi_t$ denote nuclear states and
the $B_s^{\dagger}$ and $B_t$ are the mode operators of the nuclear collective model
like in (\ref{eq:nuclearh}). Here we only label these operators by one index for
simplicity.

Substituting these operators into the matrix element of $QH_{nr}RG_0RH_{er}P$, we get
\begin{eqnarray}\label{eq:longeq}
&&\sum_r \frac{\langle q | H_{nr} | r \rangle \langle r | H_{er} | \alpha \varepsilon \rangle}{z-E_r} =
- \sum_r \sum_{abst} \sum_{\sigma\sigma'} \int d^3k \int d^3k' \frac{1}{(2\pi)^{2}} \frac{1}{k} \\
&&\times \langle q | B_s^{\dagger} B_t \int d^3 r_n \vec{j}_n \cdot\vec{\epsilon}_{\vec{k}'\sigma'} 
a_{\vec{k}'\sigma'} e^{i\vec{k}'\vec{r}_n} | r \rangle 
\nonumber \\
&&\times
\langle r | c_a^{\dagger} c_b \int d^3 r_e \phi_a^*(\vec{r}_e) \vec{\alpha}\cdot
\vec{\epsilon}_{\vec{k}\sigma}^* a_{\vec{k}\sigma}^{\dagger} e^{-i\vec{k}\vec{r}_e} \phi_b(\vec{r}_e)
| \alpha \varepsilon \rangle \,.\nonumber
\end{eqnarray}
Note that only the photon creation term of (\ref{eq:her}) and the photon
annihilation term of (\ref{eq:hnr}) contribute.  Introducing the notation
\begin{equation}
M_{ab}^e (\vec{k},\sigma) = \frac{1}{2\pi} \sqrt{\frac{c}{k}} 
\int d^3 r_e \phi_a^*(\vec{r}_e) \vec{\alpha} \cdot
\vec{\epsilon}_{\vec{k}\sigma}^* e^{-i\vec{k}\vec{r}_e} \phi_b(\vec{r}_e) \,,
\end{equation}
\begin{equation}
M_{st}^n (\vec{k}',\sigma') = - \frac{1}{2\pi} \sqrt{\frac{1}{ck}}
\int d^3 r_n \vec{j}_n \cdot
\vec{\epsilon}_{\vec{k}'\sigma'} e^{i\vec{k}'\cdot\vec{r}_n} 
\end{equation}
for the electronic and nuclear interaction matrix elements and taking into account
$\vec{k}=\vec{k}'$, $\sigma=\sigma'$, (\ref{eq:longeq}) can be condensed as
\begin{eqnarray}
&&\sum_r \frac{\langle q | H_{nr} | r \rangle \langle r | H_{er} | \alpha \varepsilon \rangle}{z-E_r} 
 =\nonumber \\
&&\sum_r \sum_{abst} \sum_{\sigma} \int d^3k 
\langle q | B_s^{\dagger} B_t a_{\vec{k}\sigma} | r \rangle
\langle r | c_a^{\dagger} c_b a_{\vec{k}\sigma}^{\dagger}         | p \rangle 
\frac{M_{ab}^e (\vec{k},\sigma) M_{st}^n (\vec{k},\sigma)}{z-E_r}\,.
\end{eqnarray}
Evaluating the above expression at $z=\varepsilon_b^e + \varepsilon_t^n$, which
is equal to the sum of the initial state electron and nuclear energies, only the
$r$ state for which $E_r = \varepsilon_a^e + \varepsilon_t^n + ck$ holds has to
be considered. Taking into account the property
\begin{equation}
\sum_{\sigma} (\vec{a}  \cdot\vec{\epsilon}_{\vec{k}\sigma})
(\vec{b} \cdot \vec{\epsilon}_{\vec{k}\sigma}) = \vec{a}\cdot  \vec{b} -
\frac{(\vec{a}\cdot  \vec{k})(\vec{b}\cdot  \vec{k})}{k^2}
\end{equation}
of the transversal polarization vectors that holds for any pair of vectors $\vec{a}$
and $\vec{b}$, we arrive to
\begin{equation}\label{eq:abst}
\sum_r \frac{\langle q | H_{nr} | r \rangle \langle r | H_{er} | \alpha \varepsilon \rangle}{z-E_r} =
\sum_{abst} \int d^3k 
\langle q | B_s^{\dagger} B_t c_a^{\dagger} c_b | p \rangle 
\frac{M_{abst} (\vec{k}) }{\varepsilon_b^e - \varepsilon_a^e - ck}\,.
\end{equation}
$M_{abst}(\vec{k})$ denotes the product of electronic and nuclear matrix elements
summed over the polarization directions
\begin{eqnarray}
&&M_{abst}(\vec{k}) = \sum_{\sigma} M_{ab}^e (\vec{k},\sigma) M_{st}^n (\vec{k},\sigma) = 
- \frac{1}{(2\pi)^2} \frac{1}{k} 
\int d^3 r_e \phi_a^*(\vec{r}_e)
\nonumber \\ &&\times
\int d^3 r_n e^{i\vec{k}\cdot(\vec{r}_n - \vec{r}_e)}
\left( \vec{j}_n  \cdot\vec{\alpha} - \frac{(\vec{j}_n \cdot\vec{k})(\vec{\alpha}\cdot \vec{k})}{k^2} \right)
\phi_b(\vec{r}_e) \,.\nonumber
\end{eqnarray}
In the long wavelength limit $\vec{k} \to \vec{0}$, this further simplifies to
\begin{equation}
M_{abst}(0) =
- \frac{1}{(2\pi)^{2}} \frac{1}{k} 
\int d^3 r_e \phi_a^*(\vec{r}_e)
\int d^3 r_n \vec{j}_n\cdot  \vec{\alpha}
\phi_b(\vec{r}_e) \,.
\end{equation}
Applying the identity
\begin{equation}\label{eq:principal}
\frac{1}{w + i\epsilon} = {\mathcal P} \frac{1}{w} -i\pi \delta(w)\ ,
\end{equation}
where ${\mathcal P}$ implies the principal value integration, and adopting the long
wavelength approximation, the real part of (\ref{eq:abst}) turns into
\begin{equation}
\lim_{\epsilon \to 0}\frac{1}{2} 
\sum_{abst} \int d^3k 
\langle q | B_s^{\dagger} B_t c_a^{\dagger} c_b | \alpha \varepsilon \rangle 
M_{abst} (\vec{k})
\left[
\frac{1}{\varepsilon_b^e - \varepsilon_a^e - ck + i\epsilon} +
\frac{1}{\varepsilon_b^e - \varepsilon_a^e - ck - i\epsilon} 
\right] \,.
\end{equation}
It can be rewritten as
\begin{equation}
\sum_{abst} \langle q | B_s^{\dagger} B_t c_a^{\dagger} c_b | \alpha \varepsilon \rangle 
\int d^3 r_e \phi_a^*(\vec{r}_e) 
V_{magn}(\vec{r}_e,\vec{r}_n; \varepsilon_b^e - \varepsilon_a^e) \phi_b(\vec{r}_e) 
\end{equation}
in terms of the effective magnetic potential
\begin{eqnarray}
V_{magn}(\vec{r}_e;\varepsilon_b^e - \varepsilon_a^e) &=&
-\frac{1}{2}\frac{1}{(2\pi)^{2}} \frac{1}{k}
\lim_{\epsilon \to 0}
\int d^3k \int d^3 r_n \left( \vec{j}_n\cdot  \vec{\alpha} - \frac{(\vec{j}_n \cdot\vec{k})(\vec{\alpha}\cdot \vec{k})}{k^2} \right) 
\nonumber \\
&\times&
e^{i\vec{k}\cdot(\vec{r}_n - \vec{r}_e)}\left(
\frac{1}{\varepsilon_b^e - \varepsilon_a^e - ck + i\epsilon} +
\frac{1}{\varepsilon_b^e - \varepsilon_a^e - ck - i\epsilon} 
\right)  \nonumber \\
&=&-\frac{1}{(2\pi)^{2}}\frac{4\pi}{|\vec{r}_n - \vec{r}_e|}  \lim_{\epsilon \to 0}
\int_0^{\infty} dk  \int d^3 r_n \left( \vec{j}_n \cdot \vec{\alpha} - \frac{(\vec{j}_n \cdot\vec{k})(\vec{\alpha}\cdot \vec{k})}{k^2} \right) 
\nonumber \\
&\times&
 \sin(k|\vec{r}_n - \vec{r}_e|)\left(
\frac{1}{\varepsilon_b^e - \varepsilon_a^e - ck + i\epsilon} +
\frac{1}{\varepsilon_b^e - \varepsilon_a^e - ck - i\epsilon} 
\right)\ . 
\end{eqnarray}

In the long wavelength limit, the $\vec{k}$-dependent part vanishes, and
$V_{magn}(\vec{r}_e;\varepsilon_b^e~-~\varepsilon_a^e)$ turns out to be
\begin{eqnarray}
V_{magn}(\vec{r}_e;0) = - \frac{1}{c}
\vec{\alpha} \int d^3 r_n \frac{\vec{j}_n(\vec{r}_n) }{|\vec{r}_n - \vec{r}_e|} \,.
\end{eqnarray}

This is equal to the magnetic Hamiltonian in (\ref{eq:hmagn}) Q.E.D. It can also
be shown that the imaginary part associated with the Dirac delta term in
(\ref{eq:principal}) vanishes if the frequency $\varepsilon_a^e -
\varepsilon_b^e$ of the exchanged photon goes to zero.


\section{\label{B}}

A further calculation of the three electronic matrix elements that enter, in eq.\
(\ref{aA1}), the formula of the magnetic Hamiltonian, gives the following
expressions for the first term $T^+_{di,\nu}$:
\begin{eqnarray}
&&T^+_{di,\nu}=-
2i(-1)^{m_d-\frac{1}{2}}
\sqrt{\frac{(2L+1)(2l_A+1)(2l_{B'}+1)}{4\pi}}
\left(\begin{array}{ccc} L & l_A & l_{B'} \\ 0 & 0 & 0 \end{array} \right)
\left(\begin{array}{ccc} L & l_A & l_{B'} \\ \nu & m+\frac{1}{2} &  \frac{1}{2}-m_d\end{array}\right)
\nonumber \\ &&\times
C\left(l_A\ \frac{1}{2}\ j;m+\frac{1}{2}\ -\frac{1}{2}\ m\right)
C\left(l_{B'}\ \frac{1}{2}\ j_d;m_d-\frac{1}{2}\ \frac{1}{2}\ m_d\right)
 \int_0^\infty dr r^{-L+1}g_{\kappa}(r)f_{\kappa_d}(r)
\nonumber \\
&&+2i(-1)^{m_d-\frac{1}{2}}
\sqrt{\frac{(2L+1)(2l_{A'}+1)(2l_B+1)}{4\pi}}
\left(\begin{array}{ccc} L & l_{A'} & l_B \\ 0 & 0 & 0 \end{array} \right)
\left(\begin{array}{ccc} L & l_{A'} & l_B \\ \nu & \frac{1}{2}-m_d & m+\frac{1}{2} \end{array} \right)
\nonumber \\ &&\times
C\left(l_{A'}\ \frac{1}{2}\ j_d;m_d-\frac{1}{2}\ \frac{1}{2}\ m_d\right)
C\left(l_B\ \frac{1}{2}\ j;m+\frac{1}{2}\ -\frac{1}{2}\ m\right)\int_0^\infty dr r^{-L+1}g_{\kappa_d}(r)f_{\kappa}(r)
\ ,
\end{eqnarray}
for the second term $T^0_{di,\nu}$:
\begin{eqnarray}
&&T^0_{di,\nu}=-i(-1)^{m_d-\frac{1}{2}}
\sqrt{\frac{(2L+1)(2l_A+1)(2l_{B'}+1)}{4\pi}}
\left(\begin{array}{ccc} L & l_A & l_{B'} \\ 0 & 0 & 0 \end{array} \right)
\left(\begin{array}{ccc} L & l_A & l_{B'} \\ \nu & m-\frac{1}{2} & \frac{1}{2}-m_d \end{array} \right)
\nonumber \\ &&\times
C\left(l_A\ \frac{1}{2}\ j;m-\frac{1}{2}\ \frac{1}{2}\ m\right)
C\left(l_{B'}\ \frac{1}{2}\ j_d;m_d-\frac{1}{2}\ \frac{1}{2}\ m_d\right)
\int_0^\infty dr r^{-L+1}f_{\kappa_d}(r)g_{\kappa}(r)
\nonumber \\
&&+i(-1)^{m_d+\frac{1}{2}}
\sqrt{\frac{(2L+1)(2l_A+1)(2l_{B'}+1)}{4\pi}}
\left(\begin{array}{ccc} L & l_A & l_{B'} \\ 0 & 0 & 0 \end{array} \right)
\left(\begin{array}{ccc} L & l_A & l_{B'} \\ \nu & m+\frac{1}{2} & -\frac{1}{2}-m_d \end{array} \right)
\nonumber \\ &&\times
C\left(l_A\ \frac{1}{2}\ j;m+\frac{1}{2}\ -\frac{1}{2}\ m\right)
C\left(l_{B'}\ \frac{1}{2}\ j_d;m_d+\frac{1}{2}\ -\frac{1}{2}\ m_d\right)
\int_0^\infty dr r^{-L+1}f_{\kappa_d}(r)g_{\kappa}(r)
\nonumber \\
&&+i(-1)^{m_d-\frac{1}{2}}
\sqrt{\frac{(2L+1)(2l_{A'}+1)(2l_B+1)}{4\pi}}
\left(\begin{array}{ccc} L & l_{A'} & l_B \\ 0 & 0 & 0 \end{array} \right)
\left(\begin{array}{ccc} L & l_{A'} & l_B \\ \nu & -\frac{1}{2}-m_d & m+\frac{1}{2} \end{array} \right)
\nonumber \\&&\times
C\left(l_{A'}\ \frac{1}{2}\ j_d;m_d+\frac{1}{2}\ -\frac{1}{2}\ m_d\right)
C\left(l_B\ \frac{1}{2}\ j;m+\frac{1}{2}\ -\frac{1}{2}\ m\right)
\int_0^\infty dr r^{-L+1}f_{\kappa}(r)g_{\kappa_d}(r)
\nonumber \\
&&+i(-1)^{m_d-\frac{1}{2}}
\sqrt{\frac{(2L+1)(2l_{A'}+1)(2l_B+1)}{4\pi}}
\left(\begin{array}{ccc} L & l_{A'} & l_B \\ 0 & 0 & 0 \end{array} \right)
\left(\begin{array}{ccc} L & l_{A'} & l_B \\ \nu & \frac{1}{2}-m_d & m-\frac{1}{2} \end{array} \right)
\nonumber \\&&\times
C\left(l_{A'}\ \frac{1}{2}\ j_d;m_d-\frac{1}{2}\ \frac{1}{2}\ m_d\right)
C\left(l_B\ \frac{1}{2}\ j;m-\frac{1}{2}\ \frac{1}{2}\ m\right)
\int_0^\infty dr r^{-L+1}f_{\kappa}(r)g_{\kappa_d}(r)\ ,
\end{eqnarray}
and for the third term $T^-_{di,\nu}$:
\begin{eqnarray}
&&T^-_{di,\nu}=-2i(-1)^{m_d+\frac{1}{2}}
\sqrt{\frac{(2L+1)(2l_A+1)(2l_{B'}+1)}{4\pi}}
\left(\begin{array}{ccc} L & l_A & l_{B'} \\ 0 & 0 & 0 \end{array} \right)
\left(\begin{array}{ccc} L & l_A & l_{B'} \\ \nu & m -\frac{1}{2} & -m_d-\frac{1}{2} \end{array} \right)
\nonumber \\&&\times
C\left(l_A\ \frac{1}{2}\ j;m-\frac{1}{2}\ \frac{1}{2}\ m\right)
C\left(l_{B'}\ \frac{1}{2}\ j_d;m_d+\frac{1}{2}\ -\frac{1}{2}\ m_d\right)
\int_0^\infty drr^{-L+1}
f_{\kappa_d}(r)g_{\kappa}(r)
 \nonumber \\
&&+2i(-1)^{m_d+\frac{1}{2}}
\sqrt{\frac{(2L+1)(2l_{A'}+1)(2l_B+1)}{4\pi}}
\left(\begin{array}{ccc} L & l_{A'} & l_B \\ 0 & 0 & 0 \end{array} \right)
\left(\begin{array}{ccc} L & l_{A'} & l_B \\ \nu & -m_d -\frac{1}{2} & m-\frac{1}{2} \end{array} \right)
\nonumber \\&&\times
C\left(l_{A'}\ \frac{1}{2}\ j_d;m_d+\frac{1}{2}\ -\frac{1}{2}\ m_d\right)
C\left(l_B\ \frac{1}{2}\ j;m-\frac{1}{2}\ \frac{1}{2}\ m\right)
\int_0^\infty drr^{-L+1}
g_{\kappa_d}(r)f_{\kappa}(r)\ .
\end{eqnarray}
Here $l_A$ and $l_B$ are the orbital quantum numbers for the upper and lower two
component spinors of the initial continuum wave function. For the wave function
of the final bound state the prime indices $l_{A'}$ and $l_{B'}$ are used. For a
given value $\kappa$, the following relations hold:
\begin{equation}
j=|\kappa|-\frac{1}{2}\nonumber \ ,
\end{equation}
\begin{equation}
l_A=\left \{\begin{array}{cc}\kappa\ &\textrm{if}\ \kappa>0\ ,\nonumber \\
|\kappa|-1\ & \textrm{if}\ \kappa<0\ ,\end{array}\right.
\end{equation}
\begin{equation}
l_B=\left \{\begin{array}{cc}\kappa-1 & \textrm{if}\ \kappa>0\ ,\\
|\kappa| & \textrm{if}\ \kappa<0\,.\end{array}\right. 
\end{equation}
\bibliography{pra}

\begin{thebibliography}{33}
\expandafter\ifx\csname natexlab\endcsname\relax\def\natexlab#1{#1}\fi
\expandafter\ifx\csname bibnamefont\endcsname\relax
  \def\bibnamefont#1{#1}\fi
\expandafter\ifx\csname bibfnamefont\endcsname\relax
  \def\bibfnamefont#1{#1}\fi
\expandafter\ifx\csname citenamefont\endcsname\relax
  \def\citenamefont#1{#1}\fi
\expandafter\ifx\csname url\endcsname\relax
  \def\url#1{\texttt{#1}}\fi
\expandafter\ifx\csname urlprefix\endcsname\relax\def\urlprefix{URL }\fi
\providecommand{\bibinfo}[2]{#2}
\providecommand{\eprint}[2][]{\url{#2}}

\bibitem[{\citenamefont{Goldanskii and Namiot}(1976)}]{Goldanskii}
\bibinfo{author}{\bibfnamefont{V.}~\bibnamefont{Goldanskii}} \bibnamefont{and}
  \bibinfo{author}{\bibfnamefont{V.~A.} \bibnamefont{Namiot}},
  \bibinfo{journal}{Phys.\ Lett.} \textbf{\bibinfo{volume}{62B}},
  \bibinfo{pages}{393} (\bibinfo{year}{1976}).

\bibitem[{\citenamefont{Harston and Chemin}(1999)}]{Harston}
\bibinfo{author}{\bibfnamefont{M.}~\bibnamefont{Harston}} \bibnamefont{and}
  \bibinfo{author}{\bibfnamefont{J.}~\bibnamefont{Chemin}},
  \bibinfo{journal}{Phys.\ Rev.\ C} \textbf{\bibinfo{volume}{59}},
  \bibinfo{pages}{2462} (\bibinfo{year}{1999}).

\bibitem[{\citenamefont{Cue et~al.}(1989)\citenamefont{Cue, Poizat, and
  Remillieux}}]{Cue}
\bibinfo{author}{\bibfnamefont{N.}~\bibnamefont{Cue}},
  \bibinfo{author}{\bibfnamefont{J.-C.} \bibnamefont{Poizat}},
  \bibnamefont{and}
  \bibinfo{author}{\bibfnamefont{J.}~\bibnamefont{Remillieux}},
  \bibinfo{journal}{Eurphys.\ Lett.} \textbf{\bibinfo{volume}{8}},
  \bibinfo{pages}{19} (\bibinfo{year}{1989}).

\bibitem[{\citenamefont{Kimball et~al.}(1991)\citenamefont{Kimball, Bittle, and
  Cue}}]{Kimball1}
\bibinfo{author}{\bibfnamefont{J.}~\bibnamefont{Kimball}},
  \bibinfo{author}{\bibfnamefont{D.}~\bibnamefont{Bittle}}, \bibnamefont{and}
  \bibinfo{author}{\bibfnamefont{N.}~\bibnamefont{Cue}},
  \bibinfo{journal}{Phys.\ Lett.} \textbf{\bibinfo{volume}{152}},
  \bibinfo{pages}{367} (\bibinfo{year}{1991}).

\bibitem[{\citenamefont{Yuan and Kimball}(1993)}]{Kimball2}
\bibinfo{author}{\bibfnamefont{Z.-S.} \bibnamefont{Yuan}} \bibnamefont{and}
  \bibinfo{author}{\bibfnamefont{J.}~\bibnamefont{Kimball}},
  \bibinfo{journal}{Phys.\ Rev.\ C} \textbf{\bibinfo{volume}{47}},
  \bibinfo{pages}{323} (\bibinfo{year}{1993}).

\bibitem[{\citenamefont{Band and Trzhaskovskaya}(1993)}]{Band}
\bibinfo{author}{\bibfnamefont{I.}~\bibnamefont{Band}} \bibnamefont{and}
  \bibinfo{author}{\bibfnamefont{M.}~\bibnamefont{Trzhaskovskaya}},
  \bibinfo{journal}{At.\ Dat.\ Nucl.\ Dat.\ Tabl.}
  \textbf{\bibinfo{volume}{55}}, \bibinfo{pages}{43} (\bibinfo{year}{1993}).

\bibitem[{\citenamefont{Kishimoto et~al.}(2000)\citenamefont{Kishimoto, Yoda,
  Seto, Kobayashi, Kitao, Haruki, Kawauchi, Fukutani, and Okano}}]{Kishimoto}
\bibinfo{author}{\bibfnamefont{S.}~\bibnamefont{Kishimoto}},
  \bibinfo{author}{\bibfnamefont{Y.}~\bibnamefont{Yoda}},
  \bibinfo{author}{\bibfnamefont{M.}~\bibnamefont{Seto}},
  \bibinfo{author}{\bibfnamefont{Y.}~\bibnamefont{Kobayashi}},
  \bibinfo{author}{\bibfnamefont{S.}~\bibnamefont{Kitao}},
  \bibinfo{author}{\bibfnamefont{R.}~\bibnamefont{Haruki}},
  \bibinfo{author}{\bibfnamefont{T.}~\bibnamefont{Kawauchi}},
  \bibinfo{author}{\bibfnamefont{K.}~\bibnamefont{Fukutani}}, \bibnamefont{and}
  \bibinfo{author}{\bibfnamefont{T.}~\bibnamefont{Okano}},
  \bibinfo{journal}{Phys.\ Rev.\ Lett.} \textbf{\bibinfo{volume}{85}},
  \bibinfo{pages}{1831} (\bibinfo{year}{2000}).

\bibitem[{\citenamefont{Carreyre et~al.}(2000)\citenamefont{Carreyre, Harston,
  Aiche, Bourgine, Chemin, Claverie, Goudour, Scheurer, Attallah, Bogaert
  et~al.}}]{Carreyre}
\bibinfo{author}{\bibfnamefont{T.}~\bibnamefont{Carreyre}},
  \bibinfo{author}{\bibfnamefont{M.~R.} \bibnamefont{Harston}},
  \bibinfo{author}{\bibfnamefont{M.}~\bibnamefont{Aiche}},
  \bibinfo{author}{\bibfnamefont{F.}~\bibnamefont{Bourgine}},
  \bibinfo{author}{\bibfnamefont{J.~F.} \bibnamefont{Chemin}},
  \bibinfo{author}{\bibfnamefont{G.}~\bibnamefont{Claverie}},
  \bibinfo{author}{\bibfnamefont{J.~P.} \bibnamefont{Goudour}},
  \bibinfo{author}{\bibfnamefont{J.~N.} \bibnamefont{Scheurer}},
  \bibinfo{author}{\bibfnamefont{F.}~\bibnamefont{Attallah}},
  \bibinfo{author}{\bibfnamefont{G.}~\bibnamefont{Bogaert}},
  \bibnamefont{et~al.}, \bibinfo{journal}{Phys.\ Rev.\ C}
  \textbf{\bibinfo{volume}{62}}, \bibinfo{pages}{024311}
  (\bibinfo{year}{2000}).

\bibitem[{\citenamefont{Wolf et~al.}(2000)\citenamefont{Wolf, Gwinner,
  Linkemann, Saghiri, Schmitt, Schwalm, Grieser, Beutelspacher, Bartsch,
  Brandau et~al.}}]{Wolf}
\bibinfo{author}{\bibfnamefont{A.}~\bibnamefont{Wolf}},
  \bibinfo{author}{\bibfnamefont{G.}~\bibnamefont{Gwinner}},
  \bibinfo{author}{\bibfnamefont{J.}~\bibnamefont{Linkemann}},
  \bibinfo{author}{\bibfnamefont{A.}~\bibnamefont{Saghiri}},
  \bibinfo{author}{\bibfnamefont{M.}~\bibnamefont{Schmitt}},
  \bibinfo{author}{\bibfnamefont{D.}~\bibnamefont{Schwalm}},
  \bibinfo{author}{\bibfnamefont{M.}~\bibnamefont{Grieser}},
  \bibinfo{author}{\bibfnamefont{M.}~\bibnamefont{Beutelspacher}},
  \bibinfo{author}{\bibfnamefont{T.}~\bibnamefont{Bartsch}},
  \bibinfo{author}{\bibfnamefont{C.}~\bibnamefont{Brandau}},
  \bibnamefont{et~al.}, \bibinfo{journal}{Nucl.\ Instrum.\ Meth.\ Phys.\ Res.\
  A} \textbf{\bibinfo{volume}{441}}, \bibinfo{pages}{183}
  (\bibinfo{year}{2000}).

\bibitem[{\citenamefont{M\"uller and Schippers}(2001)}]{Schippers}
\bibinfo{author}{\bibfnamefont{A.}~\bibnamefont{M\"uller}} \bibnamefont{and}
  \bibinfo{author}{\bibfnamefont{S.}~\bibnamefont{Schippers}},
  \bibinfo{journal}{ASP Conf.\ Series} \textbf{\bibinfo{volume}{247}},
  \bibinfo{pages}{53} (\bibinfo{year}{2001}).

\bibitem[{\citenamefont{Mart\'inez et~al.}(2005)\citenamefont{Mart\'inez,
  L\'opez-Urrutia, Braun, Brenner, Bruhns, Lapierre, Mironov, Orts, Tawara,
  Trinczek et~al.}}]{Antonio}
\bibinfo{author}{\bibfnamefont{A.~J.~G.} \bibnamefont{Mart\'inez}},
  \bibinfo{author}{\bibfnamefont{J.~R.~C.} \bibnamefont{L\'opez-Urrutia}},
  \bibinfo{author}{\bibfnamefont{J.}~\bibnamefont{Braun}},
  \bibinfo{author}{\bibfnamefont{G.}~\bibnamefont{Brenner}},
  \bibinfo{author}{\bibfnamefont{H.}~\bibnamefont{Bruhns}},
  \bibinfo{author}{\bibfnamefont{A.}~\bibnamefont{Lapierre}},
  \bibinfo{author}{\bibfnamefont{V.}~\bibnamefont{Mironov}},
  \bibinfo{author}{\bibfnamefont{R.~S.} \bibnamefont{Orts}},
  \bibinfo{author}{\bibfnamefont{H.}~\bibnamefont{Tawara}},
  \bibinfo{author}{\bibfnamefont{M.}~\bibnamefont{Trinczek}},
  \bibnamefont{et~al.}, \bibinfo{journal}{Phys.\ Rev.\ Lett.}
  \textbf{\bibinfo{volume}{94}}, \bibinfo{pages}{203201}
  (\bibinfo{year}{2005}).

\bibitem[{\citenamefont{Knapp}(1991)}]{Knapp}
\bibinfo{author}{\bibfnamefont{D.}~\bibnamefont{Knapp}}, \bibinfo{journal}{Z.
  Phys. D} \textbf{\bibinfo{volume}{21}}, \bibinfo{pages}{143}
  (\bibinfo{year}{1991}).

\bibitem[{\citenamefont{Zimmermann et~al.}(1997)\citenamefont{Zimmermann,
  Gr\"un, and Scheid}}]{Zimmermann}
\bibinfo{author}{\bibfnamefont{M.}~\bibnamefont{Zimmermann}},
  \bibinfo{author}{\bibfnamefont{N.}~\bibnamefont{Gr\"un}}, \bibnamefont{and}
  \bibinfo{author}{\bibfnamefont{W.}~\bibnamefont{Scheid}},
  \bibinfo{journal}{J.\ Phys.\ B} \textbf{\bibinfo{volume}{30}},
  \bibinfo{pages}{5259} (\bibinfo{year}{1997}).

\bibitem[{\citenamefont{Greiner and Maruhn}(1996)}]{Greiner}
\bibinfo{author}{\bibfnamefont{W.}~\bibnamefont{Greiner}} \bibnamefont{and}
  \bibinfo{author}{\bibfnamefont{J.}~\bibnamefont{Maruhn}},
  \emph{\bibinfo{title}{Nuclear Models}} (\bibinfo{publisher}{Springer Verlag
  Berlin Heidelberg}, \bibinfo{year}{1996}).

\bibitem[{\citenamefont{Zakowicz et~al.}(2004)\citenamefont{Zakowicz, Scheid,
  and Gr\"un}}]{Zakowicz}
\bibinfo{author}{\bibfnamefont{S.}~\bibnamefont{Zakowicz}},
  \bibinfo{author}{\bibfnamefont{W.}~\bibnamefont{Scheid}}, \bibnamefont{and}
  \bibinfo{author}{\bibfnamefont{N.}~\bibnamefont{Gr\"un}},
  \bibinfo{journal}{J.\ Phys.\ B} \textbf{\bibinfo{volume}{37}},
  \bibinfo{pages}{131} (\bibinfo{year}{2004}).

\bibitem[{\citenamefont{Eichler and Meyerhof}(1995)}]{Eichler}
\bibinfo{author}{\bibfnamefont{J.}~\bibnamefont{Eichler}} \bibnamefont{and}
  \bibinfo{author}{\bibfnamefont{W.}~\bibnamefont{Meyerhof}},
  \emph{\bibinfo{title}{Relativistic Atomic Collisions}}
  (\bibinfo{publisher}{Academic Press San Diego}, \bibinfo{year}{1995}).

\bibitem[{\citenamefont{Green and Rose}(1958)}]{Rose}
\bibinfo{author}{\bibfnamefont{T.}~\bibnamefont{Green}} \bibnamefont{and}
  \bibinfo{author}{\bibfnamefont{M.}~\bibnamefont{Rose}},
  \bibinfo{journal}{Phys.\ Rev.} \textbf{\bibinfo{volume}{110}},
  \bibinfo{pages}{105} (\bibinfo{year}{1958}).

\bibitem[{\citenamefont{Schwartz}(1955)}]{Schwartz}
\bibinfo{author}{\bibfnamefont{C.}~\bibnamefont{Schwartz}},
  \bibinfo{journal}{Phys.\ Rev.} \textbf{\bibinfo{volume}{97}},
  \bibinfo{pages}{380} (\bibinfo{year}{1955}).

\bibitem[{\citenamefont{Edmonds}(1996)}]{Edmonds}
\bibinfo{author}{\bibfnamefont{A.~R.} \bibnamefont{Edmonds}},
  \emph{\bibinfo{title}{Angular Momentum in Quantum Mechanics}}
  (\bibinfo{publisher}{Princeton University Press}, \bibinfo{year}{1996}).

\bibitem[{\citenamefont{Raman et~al.}(2001)\citenamefont{Raman, Nestor, and
  Tikkanen}}]{Raman}
\bibinfo{author}{\bibfnamefont{S.}~\bibnamefont{Raman}},
  \bibinfo{author}{\bibfnamefont{C.}~\bibnamefont{Nestor}}, \bibnamefont{and}
  \bibinfo{author}{\bibfnamefont{P.}~\bibnamefont{Tikkanen}},
  \bibinfo{journal}{At.\ Dat.\ Nucl.\ Dat.\ Tabl.}
  \textbf{\bibinfo{volume}{78}}, \bibinfo{pages}{1} (\bibinfo{year}{2001}).

\bibitem[{\citenamefont{Burrows}(1989)}]{NDS10}
\bibinfo{author}{\bibfnamefont{T.}~\bibnamefont{Burrows}},
  \bibinfo{journal}{Nucl.\ Dat.\ Sheets} \textbf{\bibinfo{volume}{56}},
  \bibinfo{pages}{313} (\bibinfo{year}{1989}).

\bibitem[{\citenamefont{Ring and Schuck}(1980)}]{Ring}
\bibinfo{author}{\bibfnamefont{P.}~\bibnamefont{Ring}} \bibnamefont{and}
  \bibinfo{author}{\bibfnamefont{P.}~\bibnamefont{Schuck}},
  \emph{\bibinfo{title}{The Nuclear Many-Body Problem}}
  (\bibinfo{publisher}{Springer Verlag New York}, \bibinfo{year}{1980}).

\bibitem[{\citenamefont{Parpia et~al.}(1996)\citenamefont{Parpia,
  Froese-Fischer, and Grant}}]{Par96}
\bibinfo{author}{\bibfnamefont{F.~A.} \bibnamefont{Parpia}},
  \bibinfo{author}{\bibfnamefont{C.}~\bibnamefont{Froese-Fischer}},
  \bibnamefont{and} \bibinfo{author}{\bibfnamefont{I.~P.} \bibnamefont{Grant}},
  \bibinfo{journal}{Comp.\ Phys.\ Comm.} \textbf{\bibinfo{volume}{94}},
  \bibinfo{pages}{249} (\bibinfo{year}{1996}).

\bibitem[{\citenamefont{Johnson and Soff}(1985)}]{Soff}
\bibinfo{author}{\bibfnamefont{W.~R.} \bibnamefont{Johnson}} \bibnamefont{and}
  \bibinfo{author}{\bibfnamefont{G.}~\bibnamefont{Soff}},
  \bibinfo{journal}{At.\ Dat.\ Nucl.\ Dat.\ Tabl.}
  \textbf{\bibinfo{volume}{33}}, \bibinfo{pages}{405} (\bibinfo{year}{1985}).

\bibitem[{\citenamefont{Firestone}(1991)}]{NDS1}
\bibinfo{author}{\bibfnamefont{R.~B.} \bibnamefont{Firestone}},
  \bibinfo{journal}{Nucl.\ Dat.\ Sheets} \textbf{\bibinfo{volume}{62}},
  \bibinfo{pages}{159} (\bibinfo{year}{1991}).

\bibitem[{\citenamefont{Hehner}(1988)}]{NDS9}
\bibinfo{author}{\bibfnamefont{R.~G.} \bibnamefont{Hehner}},
  \bibinfo{journal}{Nucl.\ Dat.\ Sheets} \textbf{\bibinfo{volume}{55}},
  \bibinfo{pages}{71} (\bibinfo{year}{1988}).

\bibitem[{\citenamefont{Shirley}(1988)}]{NDS2}
\bibinfo{author}{\bibfnamefont{V.~S.} \bibnamefont{Shirley}},
  \bibinfo{journal}{Nucl.\ Dat.\ Sheets} \textbf{\bibinfo{volume}{54}},
  \bibinfo{pages}{589} (\bibinfo{year}{1988}).

\bibitem[{\citenamefont{Peker}(1992)}]{NDS3}
\bibinfo{author}{\bibfnamefont{L.~K.} \bibnamefont{Peker}},
  \bibinfo{journal}{Nucl.\ Dat.\ Sheets} \textbf{\bibinfo{volume}{65}},
  \bibinfo{pages}{439} (\bibinfo{year}{1992}).

\bibitem[{\citenamefont{Browne}(1995)}]{NDS4}
\bibinfo{author}{\bibfnamefont{E.}~\bibnamefont{Browne}},
  \bibinfo{journal}{Nucl.\ Dat.\ Sheets} \textbf{\bibinfo{volume}{74}},
  \bibinfo{pages}{165} (\bibinfo{year}{1995}).

\bibitem[{\citenamefont{Reich}(1994)}]{NDS8}
\bibinfo{author}{\bibfnamefont{C.~W.} \bibnamefont{Reich}},
  \bibinfo{journal}{Nucl.\ Dat.\ Sheets} \textbf{\bibinfo{volume}{71}},
  \bibinfo{pages}{709} (\bibinfo{year}{1994}).

\bibitem[{\citenamefont{Huo}(1991)}]{NDS5}
\bibinfo{author}{\bibfnamefont{J.}~\bibnamefont{Huo}}, \bibinfo{journal}{Nucl.\
  Dat.\ Sheets} \textbf{\bibinfo{volume}{64}}, \bibinfo{pages}{723}
  (\bibinfo{year}{1991}).

\bibitem[{\citenamefont{Bhat}(1992)}]{NDS6}
\bibinfo{author}{\bibfnamefont{M.~R.} \bibnamefont{Bhat}},
  \bibinfo{journal}{Nucl.\ Dat.\ Sheets} \textbf{\bibinfo{volume}{67}},
  \bibinfo{pages}{195} (\bibinfo{year}{1992}).

\bibitem[{\citenamefont{Cameron and Singh}(2004)}]{NDS7}
\bibinfo{author}{\bibfnamefont{J.~A.} \bibnamefont{Cameron}} \bibnamefont{and}
  \bibinfo{author}{\bibfnamefont{B.}~\bibnamefont{Singh}},
  \bibinfo{journal}{Nucl.\ Dat.\ Sheets} \textbf{\bibinfo{volume}{102}},
  \bibinfo{pages}{293} (\bibinfo{year}{2004}).

\end{thebibliography}

\end{document}